\documentclass[BTech]{iitmdiss}
\usepackage{times}
\usepackage{t1enc}
\usepackage{graphicx}
\usepackage{epstopdf}
\usepackage{amsmath} 
\usepackage{bm} 
\usepackage{lipsum}
\usepackage{amssymb}
\usepackage{amsthm}
\newtheorem{lemma}{Lemma}[subsection]
\newtheorem{theorem}{Theorem}[section]

\newtheorem{definition}{Definition}[subsection]

\begin{document}

\title{Violating the assumption of Measurement Independence in Quantum Foundations}

\author{Indrajit Sen}

\date{April 2016}
\department{Physics}

\maketitle

\certificate

\vspace*{0.5in}

\noindent This is to certify that the project report titled {\bf Violating the assumption of Measurement Independence in Quantum Foundations}, submitted by {\bf Indrajit Sen}, 
  to the Indian Institute of Technology, Madras, in partial fulfilment of the requirements for
the award of the degree of {\bf Master of Science}, is a bona fide
record of the research work done by him under our supervision.  The
contents of this report, in full or in parts, have not been submitted
to any other Institute or University for the award of any degree or
diploma.

\vspace*{1.5in}

\begin{singlespacing}
\hspace*{-0.25in}
\parbox{10.0 in}{
\noindent {\bf Sibasish Ghosh \, \, \, \, \, \, \, \, \, \, \, \, \, \, \, \, \, \, \, \, \, \, \, \, \, \, \, \,\, \, \, \, \, \, \, \, \, \, \, \, \, \, \, \, \, \,Prabha Mandyam} \\
\noindent {Project Guide \, \, \, \, \, \, \, \, \, \, \, \, \, \, \, \, \, \, \, \, \, \, \, \, \, \, \, \,\, \, \, \, \, \, \, \, \, \, \, \, \, \, \, \, \, \, \, \, Project Guide}\\ 
\noindent {Associate Professor  \, \, \, \, \, \, \, \, \, \, \, \, \, \, \, \, \, \, \, \, \, \, \, \, \, \, \, \,\, \, \, \, \, \, \, \, \, \, \, \, \, \, Assistant Professor}\\
\noindent {Dept. of Physics \, \, \, \, \, \, \, \, \, \, \, \, \, \, \, \, \, \, \, \, \, \, \, \, \, \, \, \,\, \, \, \, \, \, \, \, \, \, \, \, \, \, \, \, \, Dept. of Physics}\\
\noindent {IMSc Chennai \, \, \, \, \, \, \, \, \, \, \, \, \, \, \, \, \, \, \, \, \, \, \, \, \, \, \, \,\, \, \, \, \, \, \, \, \, \, \, \, \, \, \, \, \, \, \, IIT Madras}\\

\vspace{15 mm}
Place: Chennai\\
Date: 25 April 2016
} 
\end{singlespacing}
\acknowledgements

I'd like to express my gratitude to Prof. Sibasish Ghosh, with whom I have discussed my project work almost every week since the past one year. I thank him for the enormous time he has given me, and for the ease with which I could discuss any doubt I had, no matter how small. I am thankful to him for the freedom with which I could pursue a controversial topic in Quantum foundations, the viability of which seemed dim at the beginning, and for guiding me throughout the project.\\

I'd like to thank Prof. Prabha Mandyam for introducing me to the world of Quantum foundations when I knew little about the people and literature in this field. I want to thank her for giving me the opportunity to do my Master's project in foundations, and for her considerable inputs in editing the thesis to its current form.\\

I'd also like to express my gratitude to Prof. MJW Hall who was extremely generous in entertaining all my doubts on measurement independence and related topics. He proof read and checked several results(right or wrong) that I had, pointed out connecting ideas that I had missed, and was in general very kind to give so much of his time to me. His revolutionary work in the past few years opened up the topic of measurement independence from apathy, and this thesis would not have been possible without his bold paper of 2010.\\

I'd like to thank my parents and my friends at IIT Madras for their wonderful company.\\

-Indrajit Sen
\pagebreak
\abstract

\noindent KEYWORDS: \hspace*{0.5em} \parbox[t]{4.4in}{ Quantum foundations, Measurement Independence}

\vspace*{24pt}
Though Quantum Mechanics is one of the most successful theories in Physics, experimentally verified to a very high degree of accuracy, there is still little consensus among physicists on a number of fundamental conceptual problems that the theory is plagued with since inception. Richard Feynman once remarked, "I think I can safely say that nobody understands quantum mechanics". Some of the founding fathers themselves- Erwin Schroedinger, Louis de Brogelie and Albert Einstein - were not satisfied with the theory\footnote[10]{Some of the problems are:\\
1. The Measurement problem: when and how does the collapse of the wavefunction occur?\\
2. The Quantum to Classical transition problem: exactly how many microscopic(quantum) objects must be collected together for the composite object to be macroscopic(classical)?\\
3. The problem of Reality: whether there exists an objective real world independent of our observations.\\
4. The problem of Determinism: whether Nature is fundamentally random.\\
5. The problem of Completeness: In its domain Quantum Mechanics is correct, but is it a complete, description of Nature?\\}.
The field of Quantum foundations strives to resolve these issues by reformulating the old discussions mathematically; proposing hidden variable theories that reproduce Quantum Mechanics; proposing experimental tests that can be used to decide between various standpoints; and by working on the deeper theory of Quantum Gravity from the perspective of foundations.\\

Measurement Independence is an assumption that has been used in such foundational arguments since the time of EPR thought experiment in 1935, and assumed in an overwhelming majority of hidden variable models of Quantum Mechanics since then. And yet, the term "measurement independence" is very recent - introduced only in 2010. Before this, it was known vaguely as the "free will assumption" or "lack of retrocausality". Compared to other assumptions in foundations, like locality or contextuality, it has received little attention. This is because most researchers still consider the assumption too natural to expect its violation. In the past few years however, a lot of work has been done in analysing the assumption. It has been quantified, and several interesting consequences of its violation have been derived.

In this thesis we study the various contexts in which the assumption is used, review theorems about hidden variable models in light of relaxing measurement independence, develop several new measurement dependent hidden variable models which have interesting properties, finding application in the foundational question of "reality of wavefunction", and in classical simulation of quantum channels.

\pagebreak

\begin{singlespace}
\tableofcontents
\thispagestyle{empty}

\listoffigures
\addcontentsline{toc}{chapter}{LIST OF FIGURES}
\end{singlespace}

\abbreviations

\noindent 
\begin{tabbing}
xxxxxxxxxxx \= xxxxxxxxxxxxxxxxxxxxxxxxxxxxxxxxxxxxxxxxxxxxxxxx \kill
\textbf{MI}   \> Measurement Independence \\
\textbf{MD} \> Measurement Dependent \\
\textbf{PI} \> Preparation Independence \\
\end{tabbing}
\pagebreak



\pagenumbering{arabic}

\chapter{Introduction}
\label{chap:intro}

\section{Does Quantum Foundations matter?}
\textit{"I am a Quantum Engineer, but on Sundays I have principles." - John Bell ( as quoted in \cite{1})}\\

Bohmian Mechanics \cite{2} is a completely deterministic, non-local, contextual hidden variable reformulation of standard Quantum Mechanics, where particles have classical trajectories. It reproduces all known experimental predictions of Quantum Mechanics, and has also been generalised to Quantum Field Theories \cite{3}. Thus, explanation of experimental phenomena does not require us to abandon our classical notions, say of determinism, to explain events on microscopic scale. Infact, there are several hidden variable theories\cite{4} \cite{5} \cite{6} which exactly reproduce the predictions of Quantum Mechanics, implying that no \textit{unique} interpretation can be drawn from the experimental data. Recently, hidden variable theories have also predicted phenomena which contradict Quantum Mechanics in early universe and around black holes \cite{7}, obviating the criticism of hidden variable theories that they contain no new Physics.\\

In light of this, it becomes important to ask whether the concepts used in standard Quantum Mechanics : \textit{randomness, wave-function collapse, lack of causal explanation}, are inevitable or whether, as suggested by Einstein \cite{8}, are indicative only of Quantum Mechanics being an incomplete and provisional theory, or whether, as is commonly held, a question of Semantics or Philosophy than Science. 

The result of such investigations  have relevance in Quantum Information, Quantum Gravity and Cosmology. Bell's theorem has applications in Quantum Cryptography\cite{9} and  Quantum random number generators\cite{10} Certain classes of hidden variable models can be used to classically simulate quantum channels and define communication complexity\cite{11}. The problem of defining time, and causal structures in Quantum Gravity have connections to foundations. York parameter as a candidate for time parameter is suggested by de Brogelie-Bohm theory\cite{12}. Indefinite causal structure, where causal relationships are dynamic as well as probabilistic, has been proposed for gravity by researchers in foundations \cite{13}. de Brogelie Bohm theory also predicts\cite{7} signature of violation of Born's rule in the cosmic microwave background, relic cosmological particles, Hawking radiation, photons with entangled partners inside black holes, neutrino oscillations and particles from very distant sources. 

\section{What is the Measurement Independence assumption?}
\textit{ "It is sometimes said that quantum theory saves free will. In the context of this paper.....free will saves quantum theory....in the sense of eliminating hidden variable alternatives." -C.H. Brans \cite{33}}\\

Measurement Independence(MI) is the condition of non-correlation between the hypothetical hidden variables and measurement choices made in an experiment. This assumption has been widely assumed in Quantum foundation literature. It is present in the EPR paper\cite{8}, in Bell's theorem \cite{15}, and in the framework of Ontological models\cite{16}. An early justification for this assumption was given by Bohr, in his reply to the EPR paper: "our freedom of handling the measuring instruments [is] characteristic of the very idea of experiment"\cite{11}. Measurement Independence is considered a very "reasonable" assumption by most researchers in Quantum Foundations\cite{17}. Not much work has been done on this assumption, though it has existed since the early days of Quantum foundation. Infact the term "Measurement Independence" itself has been coined very recently, in 2010 by MJW Hall\cite{18}. Also pointed out by Hall \cite{18} against the general complacency was that reasonableness alone is not sufficient - locality was very reasonable to Einstein\cite{19}. \\

Correlation between the hidden variables and the choice of measurement made by an experimenter can be due to several factors. Correlation can arise if measurement choices are no longer assumed to be acts of "free will" \cite{20} and thus uncorrelated with all events in their backward light-cones. The correlations may also result from retro-causality, wherein the event of making a choice in future affects the hidden variable state of the system in past \cite{21}.\\

\section{Outline of the thesis}
Chapter 2 introduces general hidden variable theories, and the necessary definitions that are required to understand the subsequent chapters. \\

Chapter 3 contains the bulk of work done as part of thesis. First we review some important theorems for ontological models \cite{16} in the light of relaxing MI, and consider if they still retain validity. Then we introduce the Brans \cite{33} model for singlet state correlations, which we generalise. We study the important properties of both the models, which leads us to a new result. Next, we introduce a MD model for qubits, and introduce a protocol to simulate quantum channels using it, noting the possible advantages of using MD models over ontological ones. We further introduce a MD model which has important properties relevant to EPR scenario, and discuss its foundational implications. Lastly, we show how a $\psi$ ontic model can be converted to epistemic by introducing measurement dependence in it.\\

Chapter 4 contains some observations about a different assumption, Preparation Independence, used in ref. \cite{23} to derive an important result. \\

Chapter 5 concludes with list of new results derived, and some questions that are left unanswered, serving as future directions for research.\\

\chapter{An introduction to hidden variable theories}
\section{Formulation of hidden variable theories}\label{one}
Consider an ensemble of identically prepared quantum systems, specified by a pure state. A general hidden variable theory specifies additional variables to each of these systems, usually different for each individual system. Thus, in a hidden variable theory, the systems that are identical at the quantum mechanical level differ at the hidden variable level, specified by different hidden variable configurations. A useful analogy is with that of classical statistical mechanics, where systems that are identical at the macroscopic level, say in a microcanonical ensemble each having the same energy, are different at the microscopic level, with different positions and momenta. 

Let us call the hidden variable $\lambda$. Since we are considering hidden variable theories in general, $\lambda$ can be anything: complex numbers, vectors, matrices etc. We next add the constraint upon the hidden variable theory that it reproduces the predictions of Quantum Mechanics, as we know empirically that Quantum Mechanics is correct(in non-relativistic domain). Since Quantum Mechanics gives probabilistic results which can be verified only statistically, that is over an ensemble of identically prepared systems, we require a hidden variable theory to reproduce the same probabilities by averaging over $\lambda$. That is, specification of $\lambda$ associated with each system in the ensemble will first determine the result of measurement on that particular system(which Quantum Mechanics does not give), and then we integrate over all systems in the ensemble. \\

Let a certain fraction of the ensemble have the hidden variable configuration $\lambda_{1}$,  a certain fraction have the hidden variable configuration $\lambda_{2}$...and so on. Thus, one can define the probability distribution of $\lambda$ over the entire ensemble. In general the distribution can depend on many different factors, but we will consider only the distribution conditioned over the ensemble chosen and the measurement chosen, as these two are the only variables controlled by an experimenter, i.e $p(\lambda||\psi\rangle, M)$ where the ensemble is specified by the ket $|\psi\rangle$ and $M=\{E_i\}$ is a collection of POVMs defining the measurement that will be performed on the ensemble. To be a valid probability distribution, it must be normalised
\begin{equation}
\int_{\Lambda} d\lambda p(\lambda||\psi\rangle, M) = 1 \label{1}
\end{equation}
where $\Lambda$ is the set of all possible $\lambda$s.\\
We do not however, impose the requirement of determinism at the hidden variable, so that specifying the hidden variable state $\lambda$ of the system does not necessarily give us a result with certainty. Let the result of measurement for each individual system in the ensemble be characterised by $p(k|\lambda, |\psi\rangle, M)$. If $p(k|\lambda, |\psi\rangle, M) =p^2(k|\lambda, |\psi\rangle, M)$ for all possible combinations of $k$, $\lambda$, $|\psi\rangle$ and $M$, the hidden variable theory is deterministic. Since upon measurement on a particular system, we are certain to get one of the possible results, we have
\begin{equation}
\sum_k p(k|\lambda, |\psi\rangle, M) = 1 \label{2}
\end{equation}

Finally, the sets $\{|\psi\rangle\}$ and $\{M\}$ for which the hidden variable theory is valid must be specified if it is not a general theory. For example hidden variable theories are often restricted to certain number of dimensions in Hilbert space or to projective measurements.\\

Given these, the probability of getting an outcome $k$ upon having prepared a pure state ensemble denoted by $|\psi\rangle$ and performing a measurement $M$, is
\begin{equation}
p( k| |\psi\rangle,M) = \int_{\Lambda} d\lambda p( k| |\psi\rangle, M, \lambda) p( \lambda| |\psi\rangle,M) \label{central}
\end{equation}
which is the sum over the entire ensemble.\\

Hence we have the following definition,
\begin{definition}
A \textbf{hidden variable reformulation of Quantum Mechanics} defines the following:\\
1. $\Lambda$ which is the set of all possible $\lambda$s, called the \textbf{ontic space}\cite{16}.\\
2. The probability distribution $p(\lambda| |\psi\rangle, M)$, called the \textbf{density function}\cite{16}, satisfying constraint \ref{1}.\\
3. The probability distribution $p(k|\lambda, |\psi\rangle, M)$, called the \textbf{response function}\cite{16}, satisfying constraint \ref{2}.\\
over a set of preparations defined by $\{|\psi\rangle\}$ and measurements $\{M\}$ such that the average probability of getting an outcome over the ensemble using relation \ref{central} matches with that predicted by Quantum Mechanics.
\end{definition}

Note that mixed states have not been considered yet, but the extension is simple. Consider a mixed state $\rho = \sum_i c_i |a_i\rangle \langle a_i| $ where $\sum_i c_i = 1$. Then, we have
\begin{align}
p( \lambda |\rho, M) & = \sum_{i} p( \lambda |\rho, |a_i\rangle, M) p( |a_i\rangle |\rho, M) \\
& = \sum_{a_i} p( \lambda |\rho, |a_i\rangle, M) c_i \\
& = \sum_{a_i} p( \lambda ||a_i\rangle, M) c_i  \label{3}
\end{align}
where, from the second line to third, we have assumed $p( \lambda |\rho, |a_i\rangle, M) = p( \lambda | |a_i\rangle, M)$, that is, the distribution of hidden variables in a pure state ensemble is independent of which $\rho$ the pure state is a part of. This was already implicitly assumed in eqn. \ref{1}, for the expression $p(\lambda||\psi\rangle, M)$ to make meaning.\\

Further, we have
\begin{align}
p( k |\lambda, \rho, M) & = \sum_{a_i} p( k |\lambda, \rho, |a_i\rangle, M) p( |a_i\rangle |\lambda, \rho, M) \\
& = \sum_{a_i} p( k |\lambda, |a_i\rangle, M) p( |a_i\rangle |\lambda, \rho, M) \label{4}
\end{align}
where in the second line we have assumed $p( k |\lambda, \rho, |a_i\rangle, M) = p( k |\lambda, |a_i\rangle, M)$. This was implicitly assumed in eqn. 
eqn. \ref{2}, that the probability of getting a result given the hidden variable description of an individual system which is part of a pure state ensemble, is independent of which $\rho$ the pure state is a part of. The expression $p( |a_i\rangle |\lambda, \rho, M)$ depends on the properties of the particular hidden variable theory.\\
 
Using eqns. \ref{3} and \ref{4}, we have
\begin{align}
p( k|\rho,M) &= \int_{\Lambda} d\lambda p( k|\rho, M, \lambda) p( \lambda| \rho,M) \label{5}\\
& = \int_{\Lambda} d\lambda \sum_{a_i, a_j} a_i p( \lambda ||a_i\rangle, M) p( k |\lambda, |a_j\rangle, M) p( |a_j\rangle |\lambda, \rho, M)
\end{align}

For the hidden variable theory to match experimental results, the LHS of eqn. \ref{5} must be equal to the value given by Quantum Mechanics.

\section{Formulation of Measurement Independence}

MI is the assumption of noncorrelation between the hidden variables $\lambda$ and the measurement $M$ chosen by the experimenter. So all expressions containing terms like $p(\lambda| |\psi\rangle, M)$ in section \ref{one} for theories satisfying this assumption should be replaced by $p(\lambda| |\psi\rangle)$. Historically, the EPR argument \cite{8} made the assumption that experimenters were "free" to choose whichever measurement to perform on their test system, regardless of the past. When John Bell derived his theorem\cite{15}, he replaced the "freedom of experimenter" assumption with non-correlation of hidden variables and measurement choices. This latter assumption is MI, and is infact stronger than the assumption of "experimenter's freedom". It also rules out retrocausality \cite{21}.

It is worth having a look at how the assumption enters the Quantum foundation literature, namely in Bell's theorem, which reveals how crucial it is to the central results in the field. Bell derived his inequalities twice - first in 1964\cite{15}, under the assumption of locality, determinism and MI, and again in 1976\cite{21}, under the assumption of local causality and MI.  In the next subsection we explain the standard Bell scenario common to both the theorems, and how the assumption of MI is formulated; for the complete derivations please refer to \cite{15}\cite{21}. \\

Finally, the formulation of MI in Ontological models framework \cite{16} which has been extensively used to prove various theorems recently \cite{23} \cite{24} \cite{25} \cite{26}, is discussed.\\

\subsection{Measurement Independence in Bell's 1964 Theorem}
For this section as well as the next, the scenario is the following: Consider two parties Alice and Bob, each possessing a particle entangled in spin singlet state $|\psi\rangle_{singlet} = \frac{|0\rangle|1\rangle - |1\rangle |0\rangle}{\sqrt{2}}$ with the other, and spatially separated by large distances. Let Alice measure the spin of her particle along $\hat{a}$ ( $\hat{\sigma}\cdot\hat{a}$) and Bob measure the spin of his particle along $\hat{b}$ ( $\hat{\sigma}\cdot\hat{b}$), where $\hat{k}$ is a unit vector. The question Bell asks is if correlation between the measurement results generated by Alice and Bob upon repeatedly performing such measurements, can be reproduced by a hidden variable theory which satisfies some plausible assumptions. The expectation value $\langle\hat{\sigma}\cdot\hat{a} \otimes\hat{\sigma}\cdot\hat{b} \rangle$ is calculated as,

\begin{equation}
\langle \hat{\sigma}\cdot\hat{a} \otimes\hat{\sigma}\cdot\hat{b} \rangle = \int d\lambda A(\hat{a},\lambda) B(\hat{b},\lambda) \rho(\lambda) = -\hat{a}.\hat{b} \label{6}
\end{equation}
where  $A(\hat{a},\lambda), B(\hat{b},\lambda) \in \{-1,+1\}$ are the values obtained by Alice and Bob upon measurement, given that the probability distribution $\rho(\lambda)$ of the hidden variable $\lambda$ satisfies
\begin{eqnarray}
\int \rho(\lambda) d\lambda = 1 \label{7}\\
\text{where, } \lambda \in \Lambda 
\end{eqnarray}

If $\hat{a} = \hat{b}$,  $\langle \hat{\sigma}\cdot\hat{a} \otimes \hat{\sigma}\cdot\hat{b} \rangle = -1$ from eqn. \ref{6}. Adding to eqn. \ref{7}

\begin{align}
&\int d\lambda \rho(\lambda)(1 +  A(\hat{a},\lambda) B(\hat{b},\lambda) ) = 0 \\
& \Rightarrow A(\hat{a},\lambda)B(\hat{a},\lambda) = -1\\
& \Rightarrow A(\hat{a},\lambda) = - B(\hat{a},\lambda)
\end{align}

Assuming this, and that $\lambda$ and  $\hat{a}$, $\hat{b}$ are uncorrelated, we have for the general case $\hat{a} \neq \hat{b}$, as assumed in the paper:
\begin{equation}
\langle \hat{\sigma}\cdot\hat{a}\otimes \hat{\sigma}\cdot\hat{b} \rangle = -\int d\lambda A(\hat{a},\lambda) A(\hat{b},\lambda) \rho(\lambda) \
\end{equation}
If however, $\lambda$ and $\hat{a}$, $\hat{b}$ were correlated, then $ \hat{a} = \hat{b} $ would have corresponded to $\lambda \in \Lambda_d $, where $ \Lambda_d \subset \Lambda $. For a given $\lambda \in \Lambda \setminus \Lambda_d$, it will not be possible for Alice and Bob to measure spins along $\hat{a}$ direction for their particles simultaneously, and the relation $A(\hat{a},\lambda) = B(\hat{a},\lambda)$ will be ill-defined for such $\lambda$s. Further one will have to introduce a distribution function for $\lambda$ which is correlated with the measurement directions. In general,
\begin{align}
&\rho(\lambda |\hat{a}, \hat{b}) \neq \rho(\lambda | \hat{a}, \hat{a}) \label{oho}
\end{align}

\subsection{Measurement Independence in Bell's 1976 Theorem}
\begin{figure}
\graphicspath{C:\Windows\system32\config\systemprofile\Desktop} 
\includegraphics[scale=.8]{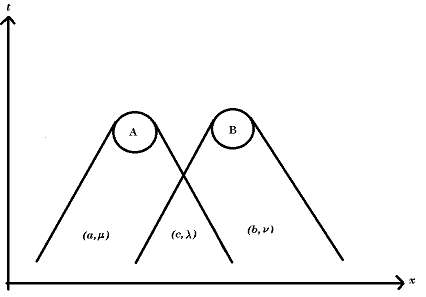}
\caption{Local hidden variables in Bell scenario}\label{figure}
\end{figure}

Bell considers local hidden variables, dividing them into non-hidden parts (a,b,c) which describe the experimental setup, and $(\mu, \upsilon, \lambda)$, the local hidden variables that are hidden(refer Fig. \ref{figure}). c lists the non-hidden variables in the overlap of the backward light cones of Alice and Bob, and a and b list non-hidden variables in the remainder of the light cones. Similarly, $\lambda$ lists the hidden variables in the overlap, and $\mu$ and $\upsilon$ list hidden variables in the remainders. Space-time regions A and B point to the measurement events taking place on 2 different instruments $M_A$ and $M_B$ respectively.
The assumption of MI is formulated as:
\begin{equation}
 p(\lambda|a,b,c) = p(\lambda|a',b,c) = p(\lambda|a,b',c) = p(\lambda|a',b',c)
\end{equation}
The above equation says that the settings of instruments $M_A$ and $M_B$, denoted by variables a and b respectively, are uncorrelated with the hidden variable $\lambda$ in the overlap of backward light cones of A and B. Assumption of non-correlation between other possible pairs of hidden and non-hidden variables, for example between $\mu$ and (a,b,c),  is not required.

\subsection{Measurement Independence in Ontological Model Framework}\label{8}
The Ontological model framework considers the experimental probability of getting an outcome $k$, given a preparation $\rho$ and measurement procedure $M$:
\begin{align}
&p( k|\rho,M) = \int p( k|\rho,M,\lambda ) p( \lambda|\rho,M) d\lambda \\
&\text{where, } \int p(\lambda| \rho, M) d\lambda = 1
\end{align}
which is simply summing up the probabilites over $\lambda $. Ontological model framework further assumes:
\begin{align}
p( k|\rho,M,\lambda ) &= p( k |M,\lambda )\label{9}\\
p( \lambda|\rho,M) &= p( \lambda|\rho)
\end{align}
The first assumption is simply to provide a framework for distinguishing $\psi$ epistemic and $\psi$ ontic theories \cite{16}( discussed in section \ref{formal} ), while the second assumption is MI.

\section{The Ontic/Epistemic distinction}
The Ontological models framework \cite{16} was introduced to tackle an unresolved debate since the early days in the subject, whether the wavefunction represents an experimenter's knowledge about the system, or is itself a property of the system. Considered a matter of philosophy by most physicists, it is a great achievement of the authors of \cite{16} to have provided a clear mathematical formulation of the issue, which was subsequently taken up in \cite{23} to give what is considered by many\cite{27} a central result in Quantum Foundations, the PBR theorem. \\

The Ontological models framework was introduced in section \ref{8}. In particular, it assumes MI and eqn. \ref{9}. Here we consider only the definition of "$\psi$ ontic" and "$\psi$ epistemic" models. 

To motivate the definitions of $\psi$ ontic and epistemic, one has to consider Einstein's argument for incompleteness of Quantum Mechanics, which is most famously put down in the EPR paper\cite{8}. However one must note Einstein himself did not write the EPR paper ( Podolsky did), and Einstein was not satisfied with the outcome, writing in a letter to Schrodinger\cite{28} shortly after the paper was published, "For reasons of language, this was written by Podolsky after many discussions. But still it has not come out as well as I really wanted; on the contrary, the main point was, so to speak, buried by the erudition.". He reproduced his own version of the argument in the same letter\cite{28}, also later in the paper 'Physics and Reality' \cite{29}, and in his autobiographical notes \cite{19}. We will concern ourselves here with Einstein's preferred argument, which will lead us more simply to the distinction between $\psi$ ontic and epistemic hidden variable theories. \textbf{Note} that in Einstein's preferred version, it is not Quantum Mechanics, but the wavefunction, which is proven an incomplete description of system.

\subsection{Einstein's Incompleteness Argument\cite{19}}\label{complete}
The following are assumed:\\
1. Two spatially separated systems have separate real states.(separability)\\
2. All interactions propagate at speed less than or equal to the speed of light.(locality)\\
3. An experimenter's choice of measurement is independent of his past.(free-will)\\
4. A system's real state is completely determined by its past.(determinism)\\

Now let us consider two particles in a spin singlet state, $|\psi\rangle = \frac{|0\rangle|1\rangle - |1\rangle |0\rangle}{\sqrt{2}}$. Let the particles be handed over to Alice and Bob, who travel away from each other, so that any communication by means of light signals between them takes a certain amount of time. From assumption 1, let the real states of Alice's particle be $\lambda_1$ and of Bob's particle be $\lambda_2$. Now consider Alice making her decision to measure her particle's spin along a certain direction \textit{after} she has separated from Bob.\\

Let Alice make a measurement along $\hat{a}$ direction. Then Bob's particle's wavefunction will collapse to an eigenket in $\hat{\sigma}\cdot{\hat{a}}$ basis. From assumption 2, as Alice is far away from Bob, her actions cannot exert a causal influence on his particle immediately. Thus, Bob's particle remains in the state $\lambda_2$.\\

From assumption 3, one can consider the case of Alice making a measurement along $\hat{a}'$ direction, without changing any event in her past. From assumption 4, as Bob's particle's past is unchanged, it still has the real state $\lambda_2$. Now however, Bob's particle's wavefunction has collapsed to an eigenket in $\hat{\sigma}\cdot{\hat{a}'}$ basis.\\

Therefore, given the 4 assumptions above, we have two different wavefunctions(eigenkets in $\hat{\sigma}\cdot{\hat{a}}$ or $\hat{\sigma}\cdot{\hat{a}'}$ basis) that describe the same real state $\lambda_2$ of Bob's particle. Thus, the wavefunction is an incomplete description of the system.\\
\textbf{Note} however the relationship, $$ \lambda \text{ described by 2 different wavefunctions } \Rightarrow \text{ wavefunction is incomplete description of system}$$ is strict. That is, $$\lambda \text{ described by only one wavefunction } \not \Rightarrow \text{ wavefunction is complete description of system.}$$ This point will become clear in the next section \ref{formal}.

\subsection{Formalization of Incompleteness in Ontological Models framework}\label{formal}
To formalise the issue of incompleteness, we have to first define the real state of a system.\\
\begin{definition}
The \textbf{real state} of the system $\lambda$ is the state, which if known, gives us the most complete knowledge of measurement results on that system. Thus, if a system has a wavefunction $|\psi\rangle$ and a measurement $M$ is performed on it, then the probability of getting the $k^{th}$ outcome given the real state $\lambda$, will have the following property:
\begin{align}
p(k|\lambda, M) = p(k|\lambda, x, M) 
\end{align}
where $x$ is any other variable. Specifically,
\begin{align}
p(k|\lambda, M) = p(k|\lambda, |\psi\rangle , M) 
\end{align}
\end{definition}
Thus, if $\lambda$ is the real state of the system, eqn. \ref{central} reduces to
\begin{align}
p( k||\psi\rangle,M) &= \int_{\Lambda} d\lambda p( k||\psi\rangle, M, \lambda) p( \lambda||\psi\rangle,M) \\
&= \int_{\Lambda} d\lambda p( k| M, \lambda) p( \lambda||\psi\rangle,M)\\
&\text{further, since ontological models assume MI,}\\
&= \int_{\Lambda} d\lambda p( k| M, \lambda) p( \lambda||\psi\rangle) \label{centrale}
\end{align} 
 
Now we are in a position to define incompleteness. From section \ref{complete}, we see that the wavefunction is an incomplete description of the system if more than one wavefunction, say $|\psi_1\rangle$ and $|\psi_2\rangle$, can correspond to the same real state $\lambda$ of the system. Infact, this tells us something stronger than incompleteness: the wavefunction describes the observer's information about the system, and not the system itself. The following discussion explains why.\\

Consider the real state of the system to be defined by $\lambda =( \lambda_1, \lambda_2, \lambda_3,....\lambda_N)$, where $\lambda_i$s are hypothetical variables which together describe the system. Now for a fixed $\lambda$, if more than one wavefunction can be used to describe $\lambda$, $|\psi\rangle$ cannot have a one to one relation with any of the $\lambda_i$s that form the description of $\lambda$. Thus the wavefunction does not describe the system itself, even incompletely. \\

It instead describes the information different observers have about the system. An analogy with classical statistical mechanics is useful here. Consider a particle in a gas defined by the microstate $(p_1,q_1)$ in the phase space, where $p_1>0$ and the Hamiltonian $H = p^2 +q^2$. Then the particle can be regarded as part of the ensemble whose momenta $p$ are positive, or as part of the ensemble(microcanonical) whose energy is $E = p_1^2 +q_1^2$. The two ensembles will be described by two different density functions $\rho$ and $\rho'$, with the crucial property that they will overlap atleast on $(p_1,q_1)$. As different observers will describe the particle by different ensembles, the density operator describes the observer's incomplete information about the system. Similarly, as the wavefunction $|\psi\rangle$ represents a pure state ensemble, assigning different wavefunctions to the same system amounts to describing the same system by different ensembles, based on incomplete information about the system.
 
Thus we have the following definitions,
\begin{definition}
If
\begin{align}
&p(\lambda||\psi_1\rangle)p(\lambda||\psi_2\rangle) > 0
\end{align}
for some $\lambda \in \Lambda$, then the model is \textbf{$\psi$ epistemic} and $|\psi\rangle$ describes an observer's information about the system.
\end{definition}

\begin{definition}
If
\begin{align}
p(\lambda||\psi_1\rangle)p(\lambda| |\psi_2\rangle) = 0 \text{ }\forall\text{ } \lambda \text{ }\in \text{ }\Lambda 
\end{align}
then, the model is \textbf{$\psi$ ontic} and $|\psi\rangle$ describes the system, either completely or incompletely. 
\end{definition}

\begin{definition}
If 
\begin{align}
p(\lambda||\psi\rangle) = \delta(\lambda - \lambda_{|\psi\rangle})\\
\end{align}
then $|\psi\rangle$ completely describes the system, and the model is called \textbf{$\psi$ ontic complete}.
\end{definition}

\begin{definition}
If
\begin{align}
&\lambda =( \lambda_1, \lambda_2, \lambda_3,....\lambda_N)\\
&\text{and} \nonumber \\
&p(\lambda||\psi\rangle) = \delta(\lambda_1 - \lambda_{|\psi\rangle})\times  p(\lambda_2, \lambda_3, \lambda_4,....\lambda_N,||\psi\rangle, \lambda_1)
\end{align}
then $|\psi\rangle$ incompletely describes the system, and the model is called \textbf{$\psi$ ontic incomplete}.
\end{definition}
\subsection{Quantifying Epistemicity in Ontological models\cite{32}}\label{epistemicity}
$\psi$ epistemic models offer an intuitive explanation of the imperfect indistinguishability of 2 non-orthogonal quantum states. To see this, we first prove a lemma. 

\begin{lemma}
For ontological models, two orthogonal kets $|\psi\rangle$ and $|\psi_{\perp}\rangle$ in Hilbert space dimension $d_N$ do not have overlapping distributions over $\Lambda$. ( even if the model is $\psi$ epistemic) \\

Consider a measurement basis $M$ consisting of the projector $|\psi_{\perp}\rangle \langle \psi_{\perp} |$.\\

From eqn. \ref{centrale}, we have 
\begin{align}
p(\psi_{\perp}| |\psi\rangle, M) &= |\langle \psi_{\perp} |\psi\rangle |^2 = 0\\
&= \int_{\Lambda} d\lambda p( \psi_{\perp}| M, \lambda) p( \lambda||\psi\rangle) = 0 \label{a}
\end{align}
and 
\begin{align}
p(\psi_{\perp}||\psi_{\perp}\rangle, M) &= |\langle \psi_{\perp} |\psi_{\perp} \rangle |^2 = 1 \\
&= \int_{\Lambda} d\lambda p( \psi_{\perp}| M, \lambda) p( \lambda||\psi_{\perp}\rangle) = 1 \label{b}
\end{align}
Eqn. \ref{a} implies the supports of $p( \psi_{\perp}| M, \lambda)$ , and $p( \lambda||\psi\rangle)$ are disjoint.( Support of a probability distribution $p(\lambda|x) ( p(y| \lambda) )$ is defined as the set $\{ \lambda \text{ } | \text{ } p(\lambda|x) ( p(y| \lambda) )\text{ }>\text{ }0\}$). But from eqn. \ref{b} we also know that the support of $p( \lambda||\psi_{\perp}\rangle)$ is a subset of support of $p( \psi_{\perp}| M, \lambda)$, as $p( \lambda||\psi_{\perp}\rangle)$ is a normalised distribution. Hence $p( \lambda| |\psi\rangle, M)$ and $p( \lambda| |\psi_{\perp}\rangle, M)$ have disjoint supports.\\
\end{lemma}
So in the case of ontological models, only non-orthogonal kets can share an overlap over $\Lambda$ space. Now suppose one is handed a system but not told the ket that describes it. Only the information that the ket is either $|\psi\rangle$ or $|\phi\rangle$ where $ 0 < |\langle \phi |\psi\rangle |^2  < 1$ is provided. To determine which ket it is, one measures it in the basis $M= \{ |\phi\rangle \langle \phi|, |\phi_{\perp}\rangle \langle \phi_{\perp}|\}$. If one gets the result $\phi$ then one says the ket was $|\phi\rangle$ and if one gets $\phi_{\perp}$ then one says the ket was $|\psi\rangle$. The process is not error-free: $|\psi\rangle$ can also give the result $\phi$, with probability $|\langle \phi |\psi\rangle |^2$. \\

For an epistemic model, there is a finite probability that the hidden variable $\lambda$ describing the system is from the overlap region between $|\psi\rangle$ and $|\phi\rangle$ in $\Lambda$ space. In that case, given only the hidden variable $\lambda$, there is no way of ascertaining which ket, $|\psi\rangle$ or $|\phi\rangle$, it came from as the result of measurement for ontological models depends only on $\lambda$. Now one can ask how much of the error probability such a model can explain. \\

Denoting by $\Lambda_{\gamma}$ the support of $p(\lambda|\gamma\rangle)$ , the probability that upon preparing a system in $|\psi\rangle$ one gets a $\lambda$ in the overlap of $\Lambda_{\phi}$ and $\Lambda_{\psi}$ is $\int_{\Lambda_{\phi}} d\lambda p(\lambda| |\psi\rangle)$.

\begin{definition}
The \textbf{degree of epistemicity} $\Omega(\psi,\phi)$ between $|\psi\rangle$ and $|\phi\rangle$ in an ontological model is defined by the equation
\begin{align}
 \int_{\Lambda_{\phi}} d\lambda p(\lambda| |\psi\rangle) = \Omega(\psi,\phi) |\langle \phi |\psi\rangle |^2
\end{align}
\end{definition}
If $\Omega(\psi,\phi) = 1$, then the probability of making an error given that the ket was actually $|\psi\rangle$ is the same as the probability that the ket $|\psi\rangle$ has a hidden variable $\lambda$ in the region $\Lambda_{\psi} \cap \Lambda_{\phi}$. Thus the model completely explains the errors in terms of overlap of distributions over $\Lambda$.

\begin{definition}
If $\Omega(\psi,\phi) = 1$ for all possible pairs $|\psi\rangle$ and $|\phi\rangle$, then the ontological model is called \textbf{maximally epistemic}.
\end{definition}
\section{Contextuality}
In section \ref{one}, we introduced the density function $p(\lambda| \rho, M)$, which depends on $\rho$. Can preparing the \textit{same} $\rho$ by different procedures lead to different $p(\lambda| \rho, M)$?\\

Consider the maximally mixed density operator in 2 dimension, 
\begin{align}
\rho &= 1/2(|0\rangle\langle 0| + |1\rangle\langle 1|) \label{10}\\
&= 1/2(|+\rangle\langle +| + |-\rangle\langle -|)\label{11}
\end{align}
where $\hat{\sigma}\cdot{\hat{z}} = (|0\rangle\langle 0| + |1\rangle\langle 1|)$ and $\hat{\sigma}\cdot{\hat{x}} = (|+\rangle\langle +| + |-\rangle\langle -|)$.\\

From eqn. \ref{10}, we see one prepare $\rho$ by having an equal number of $|0\rangle$ and $|1\rangle$ kets in the ensemble, while from eqn. \ref{11}, we see the same $\rho$ can be prepared by having an equal number of $|+\rangle$ and $|-\rangle$ kets in the ensemble. However, the second ensemble is \textit{different} from the first, though they have the same description in terms of $\rho$. 
\begin{definition}
Hidden variable models where the distribution $p(\lambda| \rho, M)$ depends on detail beyond the density operator $\rho$ are called \textbf{preparation-contextual}.
\end{definition} 
A preparation in these models is specified by not just $\rho$, but the context of preparation $S_P$, leading to distribution $p(\lambda| \rho, S_P,  M)$. For example the context here is whether one uses \ref{10} or \ref{11} to prepare one's ensemble.\\

One can easily extend the argument to pure states. Consider preparing a pure state ensemble of vertically polarized photons by two methods: \textit{a)} Passing unpolarized light through a polarizer with its transmission axis oriented vertically; and \textit{b)} Passing unpolarized light through a polarizing prism (like Wollaston prism) with its optical axis oriented such as to give us two separate beams, one horizontally polarized and another vertically, and selecting only the latter photons into our ensemble. Both methods give us the same ensemble, but the method of preparation is different. A hidden variable model which assigns different $p(\lambda| |\psi\rangle, S_P,  M)$ to the same ensemble depending on the context(method of preparation) can in principle exist, but has not been proposed so far \cite{30}.\\

There can be contexts in not only defining preparation, but also measurement. We already know from the expression $p( k|\rho, M, \lambda)$ that the response function depends on the measurement basis in general. Consider an $N$ dimensional Hilbert space where $N > 2$. Let us prepare a system in state say $\rho$ and calculate the probability of getting a measurement result corresponding to the projector $|\phi\rangle \langle \phi|$, which equals $tr(\rho |\phi\rangle \langle \phi|)$. At the hidden variable level, the probability is $p( \phi|\rho, M, \lambda)$, which must be integrated over $\Lambda$ to give $tr(\rho |\phi\rangle \langle \phi|)$.\\

However, we have several choices for our basis to perform the measurement corresponding to $|\phi\rangle \langle \phi|$. We can have the basis 
\begin{align}
&M = \{ |\phi\rangle\langle\phi|, |\phi_{\perp1}\rangle\langle\phi_{\perp1}|, |\phi_{\perp2}\rangle\langle\phi_{\perp2}|,.... |\phi_{\perp{N-1}}\rangle\langle\phi_{\perp{N-1}}| \} \\
&\text{or} \nonumber\\
&M' = \{ |\phi\rangle\langle\phi|, |\phi_{\perp1}\rangle'\langle\phi_{\perp1}|', |\phi_{\perp2}\rangle'\langle\phi_{\perp2}|',.... |\phi_{\perp{N-1}}\rangle'\langle\phi_{\perp{N-1}}|' \}
\end{align}

\begin{definition}
If $p( \phi|\rho, M, \lambda) \neq p( \phi|\rho, M', \lambda)$ in general for a hidden variable model, where $M$ contains the projector $|\phi\rangle \langle \phi|$, then the model is \textbf{measurement contextual} or \textbf{Kochen-Specker contextual}. 
\end{definition}
It means that, given $\lambda$, the probability of getting an outcome at the hidden variable level corresponding to a projector $|\phi\rangle \langle \phi|$  depends on how the projective measurement is implemented, even though the operational probability $tr(\rho |\phi\rangle \langle \phi|)$ is the same. One can similarly define contexts for POVM elements as well.\\

\chapter{Consequences of violating Measurement Independence}
In this chapter we study the consequences of violating Measurement Independence to understand the assumption better. We begin by considering several theorems proved in the ontological models framework and see if they still remain valid in the MD case.
\section{Validity of theorems for ontological models in Measurement Dependent case}\label{val}
For the case of ontological models, we have a number of results that severely constrain epistemicity. For such models, Maroney\cite{32} proved that maximal epistemicity for Hilbert space dimension $d \geq 3$ is impossible, while the epistemic explanation of indistinguishability as overlap over ontic states itself has been proven\cite{24} as arbitrarily bad for certain quantum states in $d \geq 3$. There are certain other results as well; that maximally epistemic $\Rightarrow$ Kochen-Specker noncontextual \cite{25}, and maximally epistemic $\Leftrightarrow$ Reciprocity $\cap$ Determinism \cite{30}. It is natural to ask whether these and other results can be generalized over the broader class of measurement dependent(MD) models too. In the subsections we denote by $\Lambda_\gamma$ the support of $p(\lambda||\gamma\rangle)$

\subsection{Validity of Maroney's theorem\cite{32}}
Let us first consider Maroney's result. 
\begin{theorem}
\textbf{Maroney's theorem:} For the class of ontological models, the degree of epistemicity $\Omega(\psi, \phi)$ cannot equal to 1 for arbitrary states in Hilbert space dimension greater than or equal to 3.
\end{theorem}
The argument considers three measurements $M_1, M_2 \text{ and } M_3$, and some states $|a\rangle, |b\rangle, \\
|c\rangle,|p\rangle, |m\rangle$ in Hilbert space of dimension 3. From $M_1$, it is concluded that 
\begin{equation}
\Lambda_a\cap\Lambda_p\cap\Lambda_m =\Lambda_c\cap\Lambda_p\cap\Lambda_m = \emptyset \label{1m}
\end{equation}
and from $M_2$ that 
\begin{equation}
\Lambda_b\cap\Lambda_p\cap\Lambda_m = \emptyset \label{2m}
\end{equation}
Both these results are combined to yield 
\begin{equation}
(\Lambda_a\cup\Lambda_b\cup\Lambda_c)\cap\Lambda_p\cap\Lambda_m = \emptyset \label{3m}
\end{equation} 
which is used to derive the final result. In an MD model however, the distributions over ontic space change as measurements are changed, and equations \ref{1m} and \ref{2m} cannot be combined to give \ref{3m}. The same reasoning applies to generalizing this result to dimensions greater than 3. Hence, Maroney's theorem cannot be applied to MD models.

\subsection{Validity of Barrett's theorem\cite{24}}
To state the result of Barrett \textit{et al} \cite{24}, we first need to define a few notions:
\begin{definition}
The \textbf{classical overlap} between two states $|\psi\rangle$ and $|\phi\rangle$ is defined as 
\begin{equation}
w_C = 1 - 1/2 \int_{\Lambda} d\lambda|p(\lambda|\psi\rangle) - p(\lambda| |\phi\rangle)|
\end{equation}
\end{definition}

\begin{definition}
The \textbf{quantum overlap} between two states $|\psi\rangle$ and $|\phi\rangle$ is defined as 
\begin{equation}
w_Q = 1 - \sqrt{1 - |\langle \psi|\phi \rangle|^2}
\end{equation} 
\end{definition}

\begin{theorem}
\textbf{Barrett's theorem}: No maximally epistemic ontological model can reproduce the quantum predictions for a system of dimension $d \geq 4$.\\
Moreover, as the dimension of Hilbert space $d \rightarrow \infty$, the ratio of classical over quantum overlap will tend to zero for atleast some pairs of quantum states.
\end{theorem}

The argument considers the $d + 1$ mutually unbiased orthonormal bases of a $d( \geq 4)$-dimensional Hilbert space \cite{36}, of which $|c\rangle$ is an element of one such basis, and the other bases are $\{|e_i^{\gamma}\rangle\}$, where $i, \gamma \in \{1,2,....d\}$ ($\gamma$ ranges over the bases and $i$ over the elements). From PP-incompatibility\cite{37} of $\{|c\rangle, |e_i^{\alpha}\rangle, |e_j^{\beta}\rangle\}$, a measurement $M$ having outcomes $\{f_1,f_2,f_3,f_4\}$ with the following properties is considered:
\begin{align}
&\int_{\Lambda_{e_i^{\alpha}}} p(f_1|\lambda)p(\lambda||e_i^{\alpha}\rangle)d\lambda=0\\
&\int_{\Lambda_{e_j^{\beta}}} p(f_2|\lambda)p(\lambda||e_j^{\beta}\rangle)d\lambda=0\\
&\int_{\Lambda_c} p(f_3|\lambda)p(\lambda||c\rangle)d\lambda=0\\
&\int_{\Lambda_{e_i^{\alpha}}} p(f_4|\lambda)p(\lambda| |e_i^{\alpha}\rangle)d\lambda = \int_{\Lambda_{e_j^{\beta}}} p(f_4|\lambda)p(\lambda||e_j^{\beta}\rangle)d\lambda =  \int_{\Lambda_c} p(f_4|\lambda)p(\lambda||c\rangle)d\lambda=0
\end{align}
From $M$, it is concluded that 
\begin{equation}
\Lambda_{e_i^{\alpha}} \cap \Lambda_{e_j^{\beta}} \cap \Lambda_c =\emptyset \label{4b}
\end{equation} 
It is then further assumed that, $|e_i^{\alpha}\rangle$ and $|e_j^{\alpha}\rangle$ being orthogonal,
\begin{equation}
\Lambda_{e_i^{\alpha}} \cap \Lambda_{e_j^{\alpha}}=\emptyset \label{5b}
\end{equation}
The final result is derived using both eqn. \ref{4b} and \ref{5b}. In a MD model however, eqn. \ref{5b} would actually correspond to a measurement $M' \neq M$, where the outcomes are $\{e_i^{\alpha}, e_j^{\alpha}...\}$. Two orthogonal states \textit{can} have overlapping supports in such a model, depending on the measurement being performed (see \ref{definition} for full discussion). So, Barrett \textit{et al's} result also fails to be applicable.
\subsection{Validity of Leifer-Maroney's results\cite{25}}
Ref. \cite{25} contains the following theorem: \\
\begin{theorem}
The following are true for ontological models in all dimensions of Hilbert space:\\
(i) Maximally $\psi$-epistemic $\Rightarrow$ Kochen-Specker noncontextual $\cap$ Determinism\\
(ii) Preparation noncontextual $\Rightarrow$ Maximally $\psi$-epistemic. \\
the relationship strict for both.
\end{theorem}
To check their validity in MD models, we must first generalize the notion of maximal epistemicity appropriately.\\
 
\begin{definition} \label{definition}
The \textbf{degree of epistemicity} $\Omega_M(\psi,\phi)$ between $ |\psi\rangle$ and $|\phi\rangle$, states in Hilbert space $d_N$, is defined by
\begin{equation}
\int_{\Lambda_{\phi|M}} d\lambda p( \lambda| |\psi\rangle , M)= \Omega_M(\phi, \psi) \times|\langle \psi|\phi\rangle|^2 \label{4e}
\end{equation}
when measuring both in measurement basis $M = \{ |\phi\rangle\langle\phi|, |\phi_{\perp1}\rangle\langle\phi_{\perp1}|
 , |\phi_{\perp2}\rangle\langle\phi_{\perp2}|\\,... |\phi_{\perp{N-1}}\rangle\langle\phi_{\perp{N-1}}| \}$ and where $\Lambda_{\phi|M}$ is the support of $p(\lambda| |\phi\rangle, M)$. 
\end{definition}

While generalizing this notion, it is important to consider the appropriate measurement $M$. Two orthogonal quantum states may have finite overlap over ontic space when $M$ does not contain a projector corresponding to either of the states, but nothing can be inferred from this overlap about indistinguishability. For a maximally epistemic model, $\Omega_M(\phi, \psi) = 1$ for arbitrary $|\psi\rangle$ and $|\phi\rangle$ for all $M$ that contain projector of one of them. 
\subsubsection{Validity of relation i)} 
We prove the following,
\begin{theorem}
For a measurement dependent hidden variable model, 
\begin{align}
&\text{Maximally  $\psi$ epistemic $\Rightarrow$ Determinism} \label{6l}\\
&\text{Maximally $\psi$ epistemic $\not \Rightarrow$ Kochen-Specker noncontextual}
\end{align}
\textbf{Proof}: Maximal epistemicity for measurement dependent models means\\
$\int_{\Lambda|M} p(|\phi\rangle | \lambda, M) p(\lambda | |\psi\rangle, M) d\lambda = \int_{\Lambda_{\phi}|M} p(|\phi\rangle | \lambda, M) p(\lambda | |\psi\rangle, M) d\lambda$, which implies $p(|\phi\rangle|\lambda, M) = 0$ almost everywhere on $\Lambda|M \setminus \Lambda_{\phi}|M$, where $M$ consists of $|\phi\rangle \langle \phi |$ as one of its projectors. Thus, the model is deterministic. \\

The argument is true for any other $M'$ containing $|\phi\rangle \langle \phi |$ ( it is pointless to discuss $p(|\phi\rangle|\lambda, M') $ without such context), however, as $\Lambda_{\phi}|M \neq \Lambda_{\phi}|M'$, the model is \textit{not} measurement non-contextual in general. 
\end{theorem}
The relationship is strict in MD models also as $\psi$ ontic deterministic, MD models are possible. An example is the following:
\paragraph{A $\psi$ ontic, deterministic MD model}\label{ontic}
Let the measurement basis $M = \{ |e_i\rangle \langle e_i|\}$, where $i = 1,2....n$, for a n-dimensional Hilbert space system. Let us define,
\begin{align}
&x_0= 0 \\
&x_i = | \langle e_i |\psi\rangle | \textrm{ for }  i > 0
\end{align}

Then the probability distribution of $\lambda$ is defined to be
\begin{align}
p(\lambda | |\psi\rangle, M) = | \langle e_i |\psi\rangle |\text{ for } \lambda \in (\sum_{j=0}^{i-1} x_j, \sum_{j=0}^{i} x_j) 
\end{align}

The important thing to note here is $\lambda$ is divided into n bins, each of length $| \langle e_j |\psi\rangle | \text{ for } j = 1,2..n$. The probability density of $\lambda$ in each zone is again $| \langle e_j |\psi\rangle |$. $\lambda$ is normalized ofcourse:
\begin{align}
\int p(\lambda | ||\psi\rangle, M ) d\lambda &= \sum_{j=1}^{n} \int_{x_{j-1}}^{x_{j}}  p(\lambda | |\psi\rangle, M ) d\lambda \\
&= \sum_{j=1}^{n} | \langle e_j |\psi\rangle |^2\\
&= 1
\end{align}

The response function, which explicitly depends on $|\psi\rangle$, is defined as
\begin{equation}
p(e_i | |\psi\rangle, \lambda, M) = \Theta(\lambda - \sum_{j=0}^{i-1} x_{j}) -\Theta(\lambda - \sum_{j=0}^{i} x_j)
\end{equation}
where $\Theta$ is Heaviside Step function and $x_j(|\psi\rangle, M)$.

The model reproduces Quantum Mechanics predictions:
\begin{align}
p(e_i | |\psi\rangle,  M) &= \int p(e_i | |\psi\rangle, \lambda, M) \times p( \lambda| |\psi\rangle, M) d\lambda \\
&= \int_{\sum_{j=0}^{i-1} x_{j}}^{\sum_{j=0}^{i} x_{j}}  | \langle e_i |\psi\rangle | d\lambda\\
&=| \langle e_i |\psi\rangle |^2
\end{align}

\subsubsection{Validity of relation ii)}\label{modks1}
Now let us check (ii), by considering its contrapositive as done in \cite{25}. Consider a 2-D model not maximally epistemic, so there exist $|\psi\rangle$ and $|\phi\rangle$ such that $$\int_{\Lambda|M} p(|\phi\rangle | \lambda, M) p(\lambda | |\psi\rangle, M) d\lambda > \int_{\Lambda_{\phi}|M} p(|\phi\rangle | \lambda, M) p(\lambda | |\psi\rangle, M) d\lambda$$ where $M=\{|\phi\rangle\langle\phi|,|\phi_{\perp}\rangle \langle \phi_{\perp}|\}$. Thus there exists a set of finite measure $\Omega_M$ which is disjoint from $\Lambda_{\phi|M}$ but overlaps with $\Lambda_{\psi|M}$, such that $\lambda \in \Omega_M \Rightarrow p(|\phi\rangle| \lambda, M) >0$. \\

Now consider two preparations $\rho_1  = \frac{1}{2}( |\psi\rangle\langle\psi| + |\psi_{\perp}\rangle \langle \psi_{\perp}|)= I/2$ and $\rho_2  = \frac{1}{2}( |\phi\rangle\langle\phi| + |\phi_{\perp}\rangle \langle \phi_{\perp}|)= I/2$, with corresponding distributions $p(\lambda| \rho_1, M) = \frac{1}{2}\{p(\lambda| |\psi\rangle, M) + p(\lambda| |\psi_{\perp}\rangle, M)\}$ and $p(\lambda| \rho_2, M) = \frac{1}{2}\{p(\lambda| |\phi\rangle, M) + p(\lambda| \phi_{\perp}\rangle, M)\} $, we see that $\Lambda_{\rho_1}\cap \Omega_M$ is a finite set while $\Lambda_{\rho_2}\cap \Omega_M = \emptyset$.\\

This is because, as in the specific case of ontological models, $\Omega_M$ is disjoint from both $\Lambda_{\phi|M}$ and $\Lambda_{\phi_{\perp}|M}$, however it shares an overlap with $\Lambda_{\psi|M}$. Hence the relation (ii) holds for MD models. The relation is strict again, as maximally epistemic but preparation contextual models are possible. As an example, consider a modified MD Kochen Specker model :\\

\paragraph{Modified KS Model I}
Let the hidden variable be $\lambda' = (\lambda, \hat{\lambda})$, where $\hat{\lambda}$ is a vector on the surface of Bloch sphere, and $\lambda$ is a discrete variable taking values $ \lambda_i$ and $\lambda_{i_{\perp}}$.  $\lambda$ and $\hat{\lambda}$ are correlated. The model, where $M =\{ ||i\rangle \langle i|, |i_{\perp}\rangle \langle i_{\perp}| \}$ is defined as :
\begin{align}
&p(\lambda_{i(i_{\perp})} | |\psi\rangle, M) = 1/2 \\
&p (\hat{\lambda} | |\psi\rangle, M, \lambda_{i(i_{\perp})}) = 2/\pi \times \Theta(\hat{i}(\hat{i}_{\perp})\cdot \hat{\lambda}) \times \Theta(\hat{\psi}\cdot \hat{\lambda}) \times \hat{\psi}\cdot \hat{\lambda}\\
&p(\lambda' | |\psi\rangle, M ) = p(\hat{\lambda} | |\psi\rangle, M, \lambda_k) \times p(\lambda_k | |\psi\rangle, M) \\
&p( |l\rangle \langle l| | \lambda', M) = p( |l\rangle \langle l| | \lambda_k, M) = \delta_{lk}
\end{align}
It can be checked that the model reproduces Quantum Mechanics predictions. 
\begin{align}
p( |i\rangle \langle i|| |\psi\rangle, M) &=\\
&\int \Sigma_k p( |i\rangle \langle i| | \lambda_k, M) \times p(\hat{\lambda} | |\psi\rangle, M, \lambda_k) \times p(\lambda_k | |\psi\rangle, M)d^2\hat{\lambda}\\
& = \int\Sigma_k \delta_{ik}\times  1/\pi \times \Theta(\hat{k}\cdot \hat{\lambda}) \times \Theta(\hat{\psi}\cdot \hat{\lambda}) \times \hat{\psi}\cdot \hat{\lambda}\times  d^2\hat{\lambda}\\
& = \int 1/\pi \times \Theta(\hat{i}\cdot \hat{\lambda}) \times \Theta(\hat{\psi}\cdot \hat{\lambda}) \times \hat{\psi}\cdot \hat{\lambda} \times d^2\hat{\lambda}\\
& = |\langle \psi | i \rangle |^2
\end{align}

The model is maximally epistemic as follows: Consider $|\psi\rangle$ and $|\phi\rangle$, measured in $M=\{|\psi\rangle\langle\psi|,|\psi_{\perp}\rangle\langle\psi_{\perp}|\}$. Then $p(\lambda_{\psi_{\perp}},\hat{\lambda}||\psi\rangle, M) = 0$, and  $p(\lambda_{\psi},\hat{\lambda}||\psi\rangle, M) > 0$ on a hemisphere with $\hat{\psi}$ at its center. So, the overlap over ontic space with $|\phi\rangle$ is 
\begin{align}
&\int p(\lambda_{\psi},\hat{\lambda}||\phi\rangle, M) d^2\hat{ \lambda}\\
&=\int 1/\pi \times \Theta(\hat{\psi}\cdot \hat{\lambda}) \times \Theta(\hat{\phi}\cdot \hat{\lambda}) \times \hat{\phi}\cdot \hat{\lambda} \times d^2\hat{\lambda}\\
&=|\langle \psi | \phi \rangle |^2
\end{align}
To check preparation contextuality, consider  $\rho_1  = \frac{3}{4} |0\rangle\langle 0| + \frac{1}{4}|1\rangle \langle 1|$ and $\rho_2 = \\
\frac{1}{2}(|\pi /3, 0\rangle \langle \pi /3, 0| + |\pi/3, \pi\rangle \langle \pi/3, \pi|)$, where $|\theta, \phi\rangle = \cos(\theta/2)|0\rangle + e^{i\phi}\sin(\theta/2)|1\rangle$. Though $\rho_1=\rho_2$, it can easily be checked that for $M=\{|0\rangle\langle0|, |1\rangle\langle 1| \}$, $\Lambda_{\rho_2|M} \subset \Lambda_{\rho_1|M}$.\\
\subsection{Validity of Ballentine's result\cite{30}}
Before stating Ballentine's result, we define:
\begin{definition}
The \textbf{core} $\xi_{k|M}$ of a response function $p(k|\lambda,M)$ is defined as the set $$\{ \lambda \text{ }| \text{ } p(k|\lambda,M) = 1\}$$.
\end{definition}
\begin{definition}
An ontological model satisfies \textbf{reciprocity}, or is reciprocal, if $$\Lambda_{\psi} = \xi_{\psi|M}$$ for all $M$ that contain $|\psi\rangle \langle \psi|$ as a projector.
\end{definition}
Now Ballentine's result is;
\begin{theorem}
The following relation holds for ontological models in all dimensions of Hilbert space:
\begin{equation}
Maximally \text{ }epistemic \Leftrightarrow Determinism \cap Reciprocity \label{7ba}
\end{equation}
\end{theorem}

We first generalise the notion of reciprocity for MD models
\begin{definition}
A MD hidden variabel model is \textbf{reciprocal} if the following holds 
\begin{align}
\Lambda_{\psi|M} = \xi_{\psi|M}
\end{align}
for all $M$ that contain $|\psi\rangle \langle \psi|$ as a projector, and where $\Lambda_{\psi|M}$ is the support of the distribution $p(\lambda| |\psi\rangle, M)$.
\end{definition}

From the arguments leading to eqn. \ref{6l} we can immediately confirm the forward implication of \ref{7ba}. The converse also holds true, as all contributions to \\
$\int_{\Lambda} p(\phi|\lambda, M) p(\lambda||\psi\rangle, M) d\lambda$ must come from $\Lambda_{\phi|M}$. Hence the relation is true in MD models.\\

\subsection{Validity of Bandyopadhyay \textit{et al's} results\cite{38}}
We first define a few notions used in their paper,
\begin{definition}
Upon performing a measurement $M=\{ |\phi_{1}\rangle\langle\phi_{1}|, |\phi_{2}\rangle\langle\phi_{2}|,.... |\phi_{N}\rangle\langle\phi_{N}| \}$ on state $|\psi\rangle $, the \textbf{randomness in an ontological model} valid in $d_N$, in occurrence of $i^{th}$ result( corresponding to $|\phi_{i}\rangle\langle\phi_{i}|$ is defined as :

\begin{align}
I_O(\psi,\phi_i) = \int_{\Lambda_{r|M}} d\lambda p( \phi_i| \lambda,M) \times p(\lambda| |\psi\rangle)
\end{align}
where 
\begin{equation}
\Lambda_{r|M} = \Lambda_{\psi} \cap (S_{\phi_i|M} \setminus C_{\phi_i|M})\label{r}
\end{equation} 
and $S_{\phi_i|M}$ and $C_{\phi_i|M}$ are defined as:
\begin{align}
\lambda \in C_{\phi_i|M} \Leftrightarrow p( \phi_i| \lambda,M) =1\\
\lambda \in S_{\phi_i|M} \Leftrightarrow p( \phi_i| \lambda,M) > 0
\end{align}
\end{definition}

\begin{definition}
 Upon performing a measurement $M=\{ |\phi_{1}\rangle\langle\phi_{1}|, |\phi_{2}\rangle\langle\phi_{2}|,.... |\phi_{N}\rangle\langle\phi_{N}| \}$ on state $|\psi\rangle $, the \textbf{randomness in Quantum Mechanics} in occurrence of $i^{th}$ result( corresponding to $|\phi_{i}\rangle\langle\phi_{i}|$ is defined as :
\begin{equation}
I_Q(\psi, \phi) = |\langle \psi | \phi \rangle|^2
\end{equation}

\end{definition}
The paper contains the following theorem:
\begin{theorem}
The order of randomness of ontological reciprocal models is equal to that of Quantum Mechanics for arbitrary states and measurements in Hilbert space dimension $d \geq 3$, assuming a basis independent measure of degree of epistemicity\cite{32}.
\end{theorem}

 We define randomness for MD hidden variable models in section \ref{random}, but here only note as sufficient to prove their result as not valid for MD case, that an assumption of theirs in deriving their result is to assume the validity of Maroney's theorem. As we have seen, this cannot be maintained in MD models, and thus their argument fails.\\

In section \ref{random}, we show how a MD model can be reciprocal and have \textbf{zero} randomness.\\

Hence, we see that not all theorems related to epistemicity are valid once the assumption of measurement independence is relaxed. In particular, none of the theorems that rule out maximal epistemicity in $d \geq 3$ can be extended to cover MD models, a fact that we exploit in section \ref{omain}.

\section{The Brans model\cite{33}}

Brans replaces eqn.\ref{6}  $$ \langle\hat{\sigma}\cdot\hat{a} \otimes \hat{\sigma}\cdot\hat{b}\rangle = \int d\lambda A(\hat{a},\lambda) B(\hat{b},\lambda) \rho(\lambda) $$ with 
\begin{equation}
 \langle\hat{\sigma}\cdot\hat{a} \otimes \hat{\sigma}\cdot\hat{b}\rangle = \int d\lambda A(\hat{a},\lambda) B(\hat{b},\lambda) \rho(\lambda | \hat{a}, \hat{b})
\end{equation}
so that $\lambda$ and $\hat{a}$,$\hat{b}$ are now correlated.

A simple formulation of the Brans model is as follows. Consider $\lambda$ replaced by ($\lambda'_i$, $\lambda'_j$,$\hat{A}$, $\hat{B}$) where $\lambda'_i$, $\lambda'_j$ are the parts of hidden variable describing the  2 particles, and $\hat{A}$, $\hat{B}$ are the parts of hidden variable that determine the measurement choices. Then,
\begin{align}
&\rho(\lambda'_i, \lambda'_j, \hat{A}, \hat{B}| \hat{a}, \hat{b}) = \delta(\hat{A}-\hat{a})\times\delta(\hat{B}-\hat{b})|\langle \psi_{singlet}|(|i\rangle_{\hat{A}}\otimes  |j\rangle _{\hat{B}}) |^2 \label{o}\\
& \text{where } i,j \in \{+,-\} \text{ and $|k\rangle_{\hat{A} (\hat{B})}$ denotes an eigenstate of $\hat{\sigma}\cdot \hat{A} (\hat{B})$} \nonumber\\
&  A(\lambda,\hat{a}) = A(\lambda'_i) = i\times 1 \label{a'}\\
&B(\lambda,\hat{b}) = B(\lambda'_j) = j\times 1 \label{b'}
\end{align}

The model reproduces Quantum Mechanical correlations:

\begin{align}
\langle\hat{\sigma}\cdot\hat{a} \otimes\hat{\sigma}\cdot\hat{b}\rangle &= \int d\lambda A(\hat{a},\lambda) B(\hat{b},\lambda) \rho(\lambda | \hat{a}, \hat{b}) \\
& =\int \sum_{ij} A(\lambda'_i) B(\lambda'_j) \rho(\lambda'_i, \lambda'_j, \hat{A}, \hat{B} | \hat{a}, \hat{b})  d^2\hat{A}d^2\hat{B}\\
& = \int \sum_{ij} i.j. \delta(\hat{A} -
\hat{a}).\delta(\hat{B}-\hat{b}).|\langle \psi_{singlet}|( |i\rangle_{\hat{A}}\otimes  |j\rangle _{\hat{B}}) |^2 d^2\hat{A}d^2\hat{B} \\
& = \sum_{ij} i.j.|\langle \psi_{singlet}|( |i\rangle_{\hat{A}}\otimes  |j\rangle _{\hat{B}}) |^2  \\
& = |\langle \psi_{singlet}|( |+\rangle_{\hat{A}}\otimes  |+\rangle _{\hat{B}}) |^2 - |\langle \psi_{singlet}|( |+\rangle_{\hat{A}}\otimes  |-\rangle _{\hat{B}}) |^2 \nonumber \\
 & - |\langle \psi_{singlet}|( |-\rangle_{\hat{A}}\otimes  |+\rangle _{\hat{B}}) |^2 
+ |\langle \psi_{singlet}|( |-\rangle_{\hat{A}}\otimes  |-\rangle _{\hat{B}}) |^2
\end{align}

The model satisfies Bell's locality condition and determinism, from eqns. \ref{a'} and \ref{b'} respectively. But it does not satisfy MI from \ref{o}. Let us find out how the measurement choices $\hat{a}$ and $\hat{b}$ are correlated with the hidden variables $\lambda_i'$ and $\lambda_j'$ describing the two particles.\\
\subsection{Correlation between the particles and measurement choices}\label{new}
From eqn. \ref{o}, 
\begin{align}
\rho(\lambda'_i, \lambda'_j| \hat{a}, \hat{b}) &= \int d\hat{A}d\hat{B }\rho(\lambda'_i, \lambda'_j, \hat{A}, \hat{B}| \hat{a}, \hat{b}) \\
&= \int d\hat{A} d\hat{B } \delta(\hat{A}-\hat{a})\times\delta(\hat{B}-\hat{b})|\langle \psi_{singlet}|( |i\rangle_{\hat{A}}\otimes  |j\rangle _{\hat{B}}) |^2 \\
&= |\langle \psi_{singlet}|( |i\rangle_{\hat{a}}\otimes  |j\rangle _{\hat{b}}) |^2
\end{align}

Now let's find how each individual particle is correlated with the measurement choices,
\begin{align}
\rho(\lambda'_i| \hat{a}, \hat{b}) &= \sum_j \rho(\lambda'_i, \lambda'_j| \hat{a}, \hat{b})\\
& = \sum_j|\langle \psi_{singlet}|( |i\rangle_{\hat{a}}\otimes  |j\rangle _{\hat{b}}) |^2\\
& = tr( |\psi_{singlet}\rangle\langle \psi_{singlet}||i\rangle _{\hat{a}} \langle i|_{\hat{a}} \otimes I )
\end{align}

and similarly for $\rho(\lambda'_j| \hat{a}, \hat{b})$. We thus find that the particles are correlated \textit{only} with the local measurement choices. That is, Alice's particle's hidden variable state is correlated with her choice and has not correlation with Bob's choice of measurement, and vice versa.\\

Now we generalise this model to cover arbitrary perparations and measurements. The model was first generalised by MJW Hall for arbitrary preparations and projective measurements\cite{34} having $d_1\times d_2$ outcomes, where $d_1$ and $d_2$ are integers.

\section{A maximally epistemic model in $\mathbf{d_N}$ - The generalised Brans Model}\label{omain}
For an $N$ dimensional system in Hilbert space, measuring $\rho$ with POVM elements  $M=\{ E_1, E_2, E_3,.....E_N......E_X \}$ where $ \sum_{i=1}^X E_i = \hat{I} $, the Generalized Brans model is as follows : 
\begin{align}
&\lambda \in \{\lambda_1,\lambda_2,....,\lambda_X\} \\
&p( \lambda_j|\rho, M ) =  tr(\rho E_j)  \label{3e}\\
&p( k|\lambda_j, M ) = \delta_{kj}
\end{align}

Note that a POVM in general can have any number of measurement results, even infinite\cite{35}, so $X$ is not restricted by $N$.
The model satisfies Born rule:
\begin{align}
p( k|\rho, M) &=\sum_{j}  p( k|\lambda_j, M ) p( \lambda_j|\rho, M ) \\&= \sum_{j} \delta_{kj} tr(\rho E_j)  \\&= tr(\rho E_k)
\end{align}

\subsection{How epistemic is the model?}\label{middle}
We introduced the notion of degree of epistemicity $\Omega(\psi,\phi)$ between two kets $|\psi\rangle$ and $|\phi\rangle$ for ontological models in section \ref{epistemicity}. We also generalised the notion to measurement dependent models in \ref{definition}, which we apply to the model.\\
\begin{theorem}
The generalised Brans model is maximally epistemic in arbitrary dimensions of Hilbert space.\\
\textbf{Proof}:
Consider two pure states $|\psi\rangle$ and $|\phi\rangle$ in $d_N$, satisfying $ |\langle\phi|\psi\rangle|^2 \neq 0 $. To distinguish between $|\psi\rangle$ and $|\phi\rangle$, we measure $|\psi\rangle$ in an orthonormal basis $M = \{ |\phi\rangle\langle\phi|, |\phi_{\perp1}\rangle\langle\phi_{\perp1}|,\\
 |\phi_{\perp2}\rangle\langle\phi_{\perp2}|,.... |\phi_{\perp{N-1}}\rangle\langle\phi_{\perp{N-1}}| \}$.
Now, from eqn. \ref{3e}
\begin{align}
 p( \lambda_0 | |\phi\rangle, M ) = |\langle \phi|\phi\rangle|^2  = 1\\
p( \lambda_0 | |\psi\rangle, M ) = |\langle \psi|\phi\rangle|^2
\end{align}
From eqn. \ref{4e} replacing $\int d\lambda$ by $\sum_i$
\begin{align}
&p( \lambda_0 | |\psi\rangle, M ) = |\langle \psi|\phi\rangle|^2 \times \Omega_M(\psi,\phi) \\
&\Rightarrow \Omega_M(\psi,\phi) = 1
\end{align}
\end{theorem}
 \textbf{One may note maximal epistemicity is not possible for ontological models for $\mathbf{d_N>2}$}.\\
One may consider the model in context of a POVM measurement that does not make the error of misidentification. Consider a POVM $M$ with elements
\begin{align}
&E_1 = \frac{\sqrt{2}}{1+\sqrt{2}} \times |1\rangle \langle 1 |\\
&E_2 = \frac{\sqrt{2}}{1+\sqrt{2}} \times |-\rangle \langle - |\\
&E_3 = I - E_1 - E_2
\end{align}
designed to distinguish between $|0\rangle$ and $|+\rangle$. From eqn. \ref{3e} , we have
\begin{align}
p(\lambda_3| |0\rangle, M) =  p(\lambda_3| |+\rangle, M) > 0\\
p(\lambda_1| |0\rangle, M) =  p(\lambda_2| |+\rangle, M) = 0\\
p(\lambda_2| |0\rangle, M) =  p(\lambda_1| |+\rangle, M) > 0
\end{align}
The measurement fails to distinguish between $|0\rangle$ and $|+\rangle$ when we get result $E_3$. The model explains it by saying that each time the measurement fails, the hidden variable state was in the overlap of the two kets in $\lambda_3$.\\

\subsection{Does the model satisfy Preparation Independence\cite{23}?}\label{gandom}
The question whether a model satisfies Preparation Independence (PI) is important as the PBR theorem \cite{23} rules out all epistemic models without using the assumption of MI. However the result was proven for ontological models and the definition of PI was restricted. Here we generalise this notion for MD case first.\\

\begin{definition}
A hidden variable(not necessarily ontological) model satisfies \textbf{Preparation Independence} if the following is true for a product state $|\psi_1 \rangle \otimes |\psi_2\rangle \otimes |\psi_3\rangle...|\psi_N\rangle$, 
\begin{align}
&p(\lambda | |\psi_1 \rangle \otimes |\psi_2\rangle \otimes |\psi_3\rangle...|\psi_N\rangle, M) \\
&= p(\lambda_1,\lambda_2 ,\lambda_3...\lambda_N| |\psi_1 \rangle \otimes |\psi_2\rangle \otimes |\psi_3\rangle...|\psi_N\rangle, M) \nonumber \\
& = p(\lambda_1||\psi_1 \rangle, M)\times p(\lambda_2 ||\psi_2 \rangle, M)\times p(\lambda_3| |\psi_3 \rangle, M)....\times p(\lambda_N||\psi_N \rangle, M)
\end{align}

where $p(\lambda_i||\psi_i \rangle, M)$ denotes the distribution of $|\psi_i\rangle$ over $\Lambda$ when a measurement $M$ is being performed on a $N$ dimensional tensor product state of which $|\psi_i\rangle$ is the $i^{th}$ part.\\
\end{definition}
Now let us test whether the generalised Brans model satisfies Preparation Independence(PI).
Consider $ |\psi\rangle =|\psi_1 \rangle \otimes |\psi_2\rangle \otimes |\psi_3\rangle...|\psi_N\rangle$, where say $|\psi_i\rangle$ is a qubit, so that the entire state is a $2^n$ system. Assume $\lambda = (\lambda_1, \lambda_2,\lambda_3...\lambda_n)$, where each $\lambda_k$ takes 2 values $\lambda_k^0$ and $\lambda_k^1$, which gives us $2^n$ distinct values of $\lambda$, from $\lambda^1$ to $\lambda^{2^n}$. Let us measure it in an entangled orthonormal basis $M = \{ |\phi_{1}\rangle\langle\phi_{1}|, |\phi_{2}\rangle\langle\phi_{2}|,.... |\phi_{2^n}\rangle\langle\phi_{2^n}| \}$. If $\lambda^i= (\lambda_1^a.....\lambda_k^c...\lambda_n^d)$, then
\begin{align}
&p( \lambda^i| |\psi\rangle, M ) =  p( \lambda_1^a.....\lambda_k^c...\lambda_n^d| |\psi\rangle, M ) = |\langle \phi_i |\psi\rangle|^2\\
&\text{and}\\
&p( \lambda_k^c| |\psi\rangle_k, M ) =  \sum_{\lambda_j, j \neq k} p( \lambda_1.....\lambda_k^c...\lambda_n| |\psi\rangle, M ) \\
&= \sum_m|\langle \phi_m |\psi\rangle|^2 \text{           ($2^{n-1}$ terms)}
\end{align}
Clearly,
\begin{align}
&p( \lambda^i| |\psi\rangle, M ) \neq \nonumber \\
&p( \lambda_1^a| |\psi\rangle_k, M )\times ....p( \lambda_k^c| |\psi\rangle_k, M )\times.....p( \lambda_n^d| |\psi\rangle_k, M )
\end{align}
and hence PI is not satisfied.\\


\subsection{What is the randomness in the model?}\label{random}
In a recent result, a quantification was given for the amount of randomness contained in ontological models\cite{38}. The same question can be raised for MD case, and we generalise their notion to MD case first.\\

\begin{definition}
In a hidden variable model valid in $d_N$, if we measure in an orthonormal basis $M=\{ |\phi_{1}\rangle\langle\phi_{1}|, |\phi_{2}\rangle\langle\phi_{2}|,.... |\phi_{N}\rangle\langle\phi_{N}| \}$ on state $|\psi\rangle $, the \textbf{randomness} in occurence of $i^{th}$ result( corresponding to $|\phi_{i}\rangle\langle\phi_{i}|$) is defined as :

\begin{align}
I(\psi,\phi_i) = \int_{\Lambda_{r|M}} d\lambda p( \phi_i| \lambda,M) \times p(\lambda| |\psi\rangle , M)
\end{align}
where 
\begin{equation}
\Lambda_{r|M} = \Lambda_{\psi|M} \cap (S_{\phi_i|M} \setminus C_{\phi_i|M})\label{r1}
\end{equation} 
and $S_{\phi_i|M}$ and $C_{\phi_i|M}$ are defined as:
\begin{align}
\lambda \in C_{\phi_i|M} \Leftrightarrow p( \phi_i| \lambda,M) =1\\
\lambda \in S_{\phi_i|M} \Leftrightarrow p( \phi_i| \lambda,M) > 0
\end{align}

\end{definition}
Now let us check its randomness. As this model is completely deterministic, from eqn. \ref{r1} we find $\Lambda_{r|M} = \emptyset $. Hence there is no randomness in this model. However, the Generalized Brans model is also Reciprocal, which follows directly from eqn. \ref{7ba} which holds for MD models. This proves that reciprocal models in MD case have no restrictions on randomness unlike ontological models.

\section{Simulating Quantum Channels using Measurement Dependent models}\label{bitchplease}
A quantum channel is a communication channel through which information can be transferred. In ref. \cite{39}, Montina showed how $\psi$ epistemic ontological models can be used to derive finite communication(FC) protocols for classical simulation of quantum channels.\\

The protocol Montina describes is the following. Let the quantum channel consist of Alice choosing a state $|\psi\rangle$ and sending it to Bob, who then chooses to perform a measurement $M$ on it. Bob is unaware of $|\psi\rangle$ and Alice is unaware of $M$. A classical simulation of this process using ontological models consists of the following. Alice chooses the state $|\psi\rangle$ and generates a variable $\lambda$ according to the probability distribution $p(\lambda||\psi\rangle)$. She communicates the value of $\lambda$ to Bob, who now simulates the measurement $M$ by the probability distribution $p(k| \lambda, M)$ which gives the probability of obtaining the $k^{th}$ outcome. The simulation is exact if,
\begin{align}
p( k| |\psi\rangle,M) = \int_{\Lambda} d\lambda p( k|\lambda, M) p( \lambda||\psi\rangle)
\end{align}

This is ofcourse an ontological model. Since $\lambda$ is a continuous variable in general however, Alice needs to communicate an infinite amount of information to Bob. One can reduce this communication cost by the following procedure. Instead of Alice directly communicating the value of $\lambda$, she communicates an amount of information that allows Bob to generate $\lambda$ according the distribution $p(\lambda||\psi\rangle)$. The minimum amount of communication required per round for this is equal to the mutual information $I(\lambda: |\psi\rangle)$ between $\lambda$ and $|\psi\rangle$\cite{39}. If $N$ simulations are performed in parallel, then in the limit of large $N$, the asymptotic communication cost is strictly equal to $I(\lambda: |\psi\rangle)$ \cite{39}, where for two continuous variables $x$ and $y$,
\begin{align}
&I(x: y) = h( x) + h( y) - h(x, y)\\
&\text{and}\\
&h(x) = - \int dx p(x) log_e(p(x))
\end{align}
We will not derive these results but make use of them here.\\

For a $\psi$ ontic model however, $I(\lambda: |\psi\rangle)$ is infinite as $p( \lambda||\psi\rangle)$ contains a delta function. Hence, for the simulation to have only finite amount of communication between Alice and Bob, $\psi$ epistemic ontological models are the only choice.\\

The questions we ask here are: Can MD models too simulate quantum channels? and, can simulation by MD models offer any advantage over simulation by ontological models? We first give a MD hidden variable model for qubits and then give a protocol to use it to simulate quantum channels. Later we discuss how measurement dependent models can be more advantageous than ontological models for such simulations.

\subsection{Modified Kochen Specker model II}\label{modks2}
Let the qubit $|a\rangle$ be denoted by $\hat{a}$ on the Bloch sphere, and the measurement $M = \{ |b\rangle \langle b|, |b_{\perp}\rangle \langle b_{\perp}|\}$ be denoted by $\hat{b}$ on the Bloch sphere. Then, the model is defined by
\begin{align}
&p(\hat{\lambda} | \hat{a}, \hat{b}) = \frac{\Theta(\hat{\lambda} \cdot \hat{a}) |\hat{\lambda} \cdot \hat{b }|}{\pi}  \label{1ks}  \\
&p(\hat{k}| \hat{\lambda}) = \Theta(\hat{\lambda} \cdot \hat{k}) \label{2ks}
\end{align}
where  $\Theta$ is the Heaviside Step function, and $\hat{k} \in \{\hat{b}, -\hat{b}\}$.
The density function for $\hat{\lambda}$ is normalized, 
\begin{align}
\int p(\hat{\lambda} | \hat{a}, \hat{b}) d\hat{\lambda} = 1
\end{align}
and the model reproduces Quantum Mechanics predictions, 
\begin{align}
p(\pm \hat{b}|\hat{a},\hat{b}) &= \int \frac{\Theta(\hat{\lambda} \cdot \hat{a}) |\hat{\lambda} \cdot \hat{b }|}{\pi} \Theta(\pm \hat{\lambda} \cdot \hat{b}) d\hat{\lambda}\\
&= \frac{1 \pm\hat{a}\cdot \hat{b}}{2}\\
\end{align}
 
It can be checked that the model is maximally epistemic.
\subsection{Protocol to simulate quantum channels using modified Kochen Specker model II}
Let Alice prepare her qubit along $\hat{a}$. As she does not know what measurement $\hat{b}$ Bob will choose, she cannot generate the complete probability distribution \ref{1ks}. She instead generates a uniform distribution over the hemisphere with $\hat{a}$ at its center, $\Theta(\hat{\lambda} \cdot \hat{a})/2\pi$. She sends $\hat{\lambda}$ according to this distribution to Bob.\\

At his end Bob does not accept all the $\hat{\lambda}$s Alice is sending. He first chooses his measurement $\hat{b}$ and then attaches a weight of $2|\hat{\lambda} \cdot \hat{b}|$ to the uniform distribution sent by Alice. He picks up more $\hat{\lambda}$s from the regions $|\hat{\lambda} \cdot \hat{b}|$ is high and less from the regions where $|\hat{\lambda} \cdot \hat{b}|$ is low so as to generate the final distribution $\Theta(\hat{\lambda} \cdot \hat{a}) |\hat{\lambda} \cdot \hat{b }|/\pi$. Effectively, Bob picks up $\hat{\lambda}$s with probability distribution $|\hat{\lambda} \cdot \hat{b}|/\pi$ over the hemisphere defined by $\Theta(\hat{\lambda} \cdot \hat{a})$.\\

In this protocol, Alice does not generate the final distribution of $\hat{\lambda}$. She only has to communicate such that Bob can generate a uniform distribution over the hemisphere with $\hat{a}$ at its center. The communication required is $I(\hat{a} : \hat{\lambda})$. Let us calculate the same.\\

\subsection{Calculation of Communication cost}
Here,
\begin{align}
&p(\hat{\lambda} | \hat{a}, \hat{b}) = \frac{\Theta(\hat{\lambda} \cdot \hat{a}) |\hat{\lambda} \cdot \hat{b }|}{\pi}\\
&p(\hat{b}| \hat{\lambda}) = \Theta(\hat{\lambda} \cdot \hat{b})
\end{align}

Assuming  $p(\hat{a}) = \frac{1}{4 \pi}$, we have,
\begin{align}
h(\hat{a}) &= -\int \frac{1}{4 \pi} \log_e ( \frac{1}{4 \pi}) d\hat{a}\\
&= \log_e (4 \pi)
\end{align}
 As $\hat{\lambda}$ depends on both $\hat{a}$ and $\hat{b}$, assuming $p(\hat{b}) = \frac{1}{4 \pi}$ we first find
\begin{align}
p(\hat{\lambda}|\hat{a}) &= \int p(\hat{\lambda}|\hat{a}, \hat{b}) p(\hat{b}) d\hat{b}\\
&= \frac{1}{2 \pi}\Theta(\hat{\lambda} \cdot \hat{a})
\end{align}

and,
\begin{align}
p(\hat{\lambda}) &= \int p(\hat{\lambda}| \hat{a})p(\hat{a}) d\hat{a}\\
&=\int  \frac{1}{2 \pi}\Theta(\hat{\lambda} \cdot \hat{a}) \times \frac{1}{4 \pi} d\hat{a}\\
& = \frac{1}{4 \pi}
\end{align}
Therefore, $h(\hat{\lambda}) = h(\hat{a}) = \log_e (4 \pi)$.
Now, 
\begin{align}
h(\hat{\lambda}, \hat{a})& = - \int p(\hat{\lambda}|\hat{a}) p(\hat{a}) log_e (p(\hat{\lambda}|\hat{a})p(\hat{a})) d\hat{\lambda} d\hat{a}\\
&= -\frac{1}{8 \pi^2} \int \Theta(\hat{\lambda} \cdot \hat{a}) \log_e ( \frac{\Theta(\hat{\lambda} \cdot \hat{a})} {8\pi^2}) d\hat{\lambda}d\hat{a}\\
& = \frac{1}{8 \pi^2} \{ \log_e(8\pi^2) \int \Theta(\hat{\lambda} \cdot \hat{a}) d\hat{\lambda}d\hat{a} - \int \Theta(\hat{\lambda} \cdot \hat{a}) \log_e( \Theta(\hat{\lambda} \cdot \hat{a})) d\hat{\lambda} d\hat{a} \}\\
&= \log_e(2) \text{ nats} = 1 \text{ bit}
\end{align}

Now Alice sends information for each $\hat{\lambda}$. But Bob does not use all the $\hat{\lambda}$s Alice is sending. Hence we will have a correction factor for this. Alice sends "$2\pi$" amount of $\hat{\lambda}$s to Bob, as she generates a uniform distribution over a hemisphere. Bob finally selects $\pi$ amount of $\hat{\lambda}$, as the final distribution is  $\Theta(\hat{\lambda} \cdot \hat{a}) |\hat{\lambda} \cdot \hat{b }|/\pi$. Thus, he selects only half of the $\hat{\lambda}$s Alice is sending. As Alice sends information for all $\hat{\lambda}$s nevertheless, the communication cost involved in this protocol is twice that of calculated, \textbf{2 bits} per round.\\

We thus see that it is indeed possible to use MD models for classical simulation of quantum channels.

\subsection{Advantage of using Measurement Dependent models for simulation}
As we saw in section \ref{middle}, MD models are not constrained in their maximal epistemicity in higher dimensions of Hilbert space unlike ontological models, where maximal epistemicity is impossible for $d_N >2$. Thus, one can in principle develop protocols for classical simulation of quantum channels invloving qutrits or higher dimensional systems, by using MD maximally epistemic hidden variable models.

\section{A Measurement Dependent model that cross correlates particles and measurement choices in EPR scenario}
In section \ref{new} we saw that in the Brans model, the hidden variable state of Alice has no correlation with the measurement choice of Bob, and vice versa. It is of interest to develop a model where the particles are correlated with both measurement choices, as this has foundational implications, discussed in section \ref{ken}. First we present a model that achieves this.
\subsection{Modified Hall Model}\label{ryu}
The model is a modified version of the local, deterministic and MD model for singlet state correlations given by Hall in \cite{18}, and reformulated in \cite{34}. Here we make it separable too.\\

Let the ontic states of the two particles entangled in singlet state be denoted by $\hat{\lambda}_1$ and $\hat{\lambda}_2$; both are vectors on a unit sphere. Let the corresponding experimenters make measurements along $\hat{a}$ and $\hat{b}$ directions. Then, the measurement results are given by,
\begin{align}
&A(\hat{\lambda}_1,\hat{a}) = Sign(\hat{\lambda}_1\cdot\hat{a})\label{0+}\\
&B(\hat{\lambda}_2,\hat{b}) = Sign(\hat{\lambda}_2\cdot\hat{b})\label{0++}
\end{align}
where $Sign(x)$ is the sign function. So, the model is deterministic and local.\\

The probability distribution of the ontic states is given by,
\begin{align}
p(\hat{\lambda}_1,\hat{\lambda}_2|\hat{a},\hat{b}, |\psi\rangle_{singlet}) = \frac{1}{4\pi} \frac{1-(\hat{a}\cdot\hat{b})Sign\{(\hat{\lambda}_1\cdot \hat{a})(\hat{\lambda}_2\cdot \hat{b})\}}{ 1-(1-2\frac{\phi_{\hat{a}\hat{b}}}{\pi}) Sign\{(\hat{\lambda}_1\cdot \hat{a})(\hat{\lambda}_2\cdot \hat{b})\}} \delta(\hat{\lambda}_1 + \hat{\lambda}_2)
\end{align}
where $\phi_{\hat{a}\hat{b}}$ is the angle between $\hat{a}$ and $\hat{b}$.\\

The marginals are,
\begin{align}
p(\hat{\lambda}_1|\hat{a},\hat{b}, |\psi\rangle_{singlet}) &= \int d\hat{\lambda}_2 p(\hat{\lambda}_1,\hat{\lambda}_2|\hat{a},\hat{b}, |\psi\rangle_{singlet})\\
&= \frac{1}{4\pi} \frac{1+(\hat{a}\cdot\hat{b})Sign\{(\hat{\lambda}_1\cdot \hat{a})(\hat{\lambda}_1\cdot \hat{b})\}}{ 1+(1-2\frac{\phi_{\hat{a}\hat{b}}}{\pi}) Sign\{(\hat{\lambda}_1\cdot \hat{a})(\hat{\lambda}_1\cdot \hat{b})\}}\label{1+}
\end{align}
And similar for $p(\hat{\lambda}_2|\hat{a},\hat{b}, |\psi\rangle_{singlet})$.\\

From eqn. \ref{1+} we see that $\hat{\lambda}_1$ is correlated with measurement choices of \textit{both} experimenters. \\

That the model correctly reproduces singlet state correlations can be proven. The probability Alice gets value $x$ and Bob $y$ on measuring spins along $\hat{\sigma}\cdot{\hat{a}}$ and $\hat{\sigma}\cdot{\hat{b}}$ respectively is,

\begin{align}
&p(x,y||\psi\rangle_{singlet}, M = \hat{\sigma}\cdot{\hat{a}} \otimes \hat{\sigma}\cdot{\hat{b}}) = \int d\hat{\lambda}_1 d\hat{\lambda}_2 \delta_{x,A(\hat{\lambda}_1,\hat{a})} \delta_{y,B(\hat{\lambda}_2,\hat{b})} p(\hat{\lambda}_1,\hat{\lambda}_2|\hat{a},\hat{b}, |\psi\rangle_{singlet}) \\
& = \int d\hat{\lambda}_1 d\hat{\lambda}_2 \delta_{x,A(\hat{\lambda}_1,\hat{a})} \delta_{y,B(\hat{\lambda}_2,\hat{b})} \frac{1}{4\pi} \frac{1-(\hat{a}\cdot\hat{b})Sign\{(\hat{\lambda}_1\cdot \hat{a})(\hat{\lambda}_2\cdot \hat{b})\}}{ 1-(1-2\frac{\phi_{\hat{a}\hat{b}}}{\pi}) Sign\{(\hat{\lambda}_1\cdot \hat{a})(\hat{\lambda}_2\cdot \hat{b})\}} \delta(\hat{\lambda}_1 + \hat{\lambda}_2) \\
& = \int d\hat{\lambda}_1 \delta_{x,A(\hat{\lambda}_1,\hat{a})} \delta_{y,B(-\hat{\lambda}_1,\hat{b})} \frac{1}{4\pi} \frac{1+(\hat{a}\cdot\hat{b})Sign\{(\hat{\lambda}_1\cdot \hat{a})(\hat{\lambda}_1\cdot \hat{b})\}}{ 1+(1-2\frac{\phi_{\hat{a}\hat{b}}}{\pi}) Sign\{(\hat{\lambda}_1\cdot \hat{a})(\hat{\lambda}_1\cdot \hat{b})\}} \label{2+}
\end{align} 
where $\delta$ is the Kronecker delta function. Eqn. \ref{2+} is the same integral as in \cite{34}, thus ensuring correctness.

\subsection{Foundational implications of such a model}\label{ken}
In reference \cite{16}, the authors claim to prove a theorem that any locally causal \cite{22}, separable model that reproduces quantum mechanical predictions must be $\psi$ epistemic in EPR scenario. Their argument consists of the following assumption: Consider a frame where Alice makes the first measurement. Let her choose her measurement direction either along $\hat{a}$ or $\hat{a'}$. If the model is locally causal, then they argue that the measurement choices made by Alice, at spacelike separation from Bob, should not affect the distribution of Bob's particle's hidden variable state $\lambda_B$.
\begin{equation}
p(\lambda_{B}|\hat{a}) = p(\lambda_{B}|\hat{a'})\label{wrong}
\end{equation}

\begin{figure}
\graphicspath{C:\Windows\System32\config\systemprofile\Documents\Physics and Philosophy\Thesis} 
\includegraphics[scale=0.8]{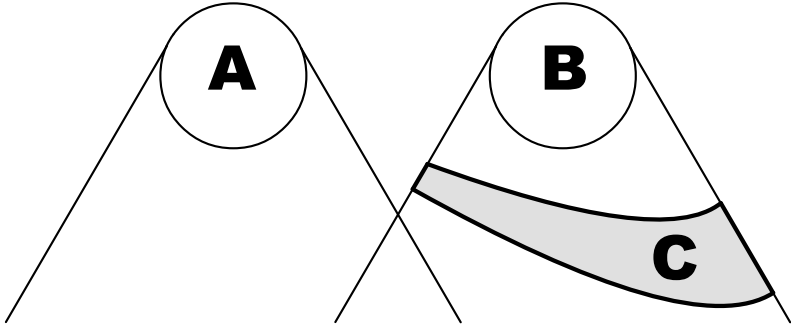}
\caption{Condition of local-causality}\label{figure1}
\end{figure}

The assumption is incorrect to the best of our understanding. Instead of being a consequence of local-causality, it is a consequence of MI. As defined in their own paper in agreement with Bell \cite{22},
\begin{definition}
Consider an event $x$ occuring at spacetime region A and an event $y$ occuring at spacetime region B, where A and B are space-like separated. Then \textbf{local-causality} is the condition that
\begin{equation}
p(x| \lambda_C, y) = p(x|y) 
\end{equation}
where $\lambda_C$ contains a \textbf{complete} specification of events in space time region C that screens off B from the intersection of backward light cones of A and B. (refer Fig. \ref{figure1})
\end{definition}
No such complete description is provided in eqn. \ref{wrong}. But the equation makes sense if one assumes MI, as the measurement choices then have correlations only with events in their future light cones.\\
Now let us consider the MD model introduced in the last section \ref{ryu}, which leads to the following theorem:

\begin{theorem}
A measurement dependent model which is local causal need not be $\psi$ epistemic.\\

\textbf{Proof}:
Consider that, in a frame where Alice makes the first measurement, she chooses her measurement direction either along $\hat{a}$ or $\hat{a'}$. Accordingly, the probability distribution of $\hat{\lambda}_2$ of Modified Hall model, 
\begin{align}
&p(\hat{\lambda}_2|\hat{a},\hat{b}, |\psi\rangle_{singlet}) = \frac{1}{4\pi} \frac{1+(\hat{a}\cdot\hat{b})Sign\{(\hat{\lambda}_2\cdot \hat{a})(\hat{\lambda}_2\cdot \hat{b})\}}{ 1+(1-2\frac{\phi_{\hat{a}\hat{b}}}{\pi}) Sign\{(\hat{\lambda}_2\cdot \hat{a})(\hat{\lambda}_2\cdot \hat{b})\}}\\
&p(\hat{\lambda}_2|\hat{a'},\hat{b}, |\psi\rangle_{singlet}) = \frac{1}{4\pi} \frac{1+(\hat{a'}\cdot\hat{b})Sign\{(\hat{\lambda}_2\cdot \hat{a'})(\hat{\lambda}_2\cdot \hat{b})\}}{ 1+(1-2\frac{\phi_{\hat{a'}\hat{b}}}{\pi}) Sign\{(\hat{\lambda}_2\cdot \hat{a'})(\hat{\lambda}_2\cdot \hat{b})\}}
\end{align}
\textit{differs}, but this does not imply a failure of local-causality. Eqns. \ref{0+} and \ref{0++} explicitly show the local nature of the model. \\
Hence, the argument that they should be same as a consequence of local-causality, cannot be held for MD models, and epistemicity cannot be derived as in \cite{16}.
\end{theorem}
\section{Can a $\mathbf{\psi}$ ontic ontological model be converted to $\mathbf{\psi}$ epistemic by introducing Measurement Dependence?}
In sections \ref{modks1} and \ref{modks2}, we modified an already maximally epistemic ontological model\cite{31} to maximally epistemic MD model in 2 ways. The question remains if one can convert a $\psi$ ontic ontological model to $\psi$ epistemic by introducing measurement dependence. Here we prove it in the affirmative by modifying the $\psi$ ontic Bell-Mermin model\cite{40} to a maximally $\psi$ epistemic MD model.

\subsection{Modified Bell Mermin model}\label{BM}
Here, the hidden variable $\lambda' = (\lambda, \hat{\lambda})$, where $\hat{\lambda}$ is a vector on the surface of Bloch sphere, and $\lambda$ is a discrete variable taking values $ \lambda_i$ and $\lambda_j$. $\lambda$ and $\hat{\lambda}$ are correlated. The model, where $M=\{ |i\rangle \langle i|, |j\rangle \langle j| \}$ is an orthonormal basis, is defined as :
\begin{align}
&p(\lambda_{k=i(j)} | |\psi\rangle, M) = 1/2 \\
&p (\hat{\lambda} | |\psi\rangle, M, \lambda_k) = 1/(2\pi) \times \Theta( \hat{k} \cdot ( \hat{\psi} + \hat{\lambda} ))\\
&p(\lambda' | |\psi\rangle, M ) = p(\hat{\lambda} | |\psi\rangle, M, \lambda_k) \times p(\lambda_k | |\psi\rangle, M) \\
&p( |l\rangle \langle l| | \lambda', M) = p( |l\rangle \langle l| | \lambda_k, M) = \delta_{lk}
\end{align}
where $\Theta$ is the step function. The model reproduces Quantum Mechanics predictions:
\begin{align}
p( |i\rangle \langle i|| |\psi\rangle, M) &= \int \sum_k p( |i\rangle \langle i| | \lambda_k, M) \times p(\hat{\lambda} | |\psi\rangle, M, \lambda_k) \times p(\lambda_k | |\psi\rangle, M)d^2\hat{\lambda}\\
& = \int\sum_k \delta_{ik}\times 1/(4\pi) \times \Theta( \hat{k} \cdot ( \hat{\psi} + \hat{\lambda} )) d^2\hat{\lambda}\\
& = \int 1/(4\pi) \times \Theta( \hat{i} \cdot ( \hat{\psi} + \hat{\lambda} )) d^2\hat{\lambda}\\
& = |\langle \psi | i \rangle |^2
\end{align}

It can be checked that the model is maximally $\psi$ epistemic, as follows. Consider two states $|i\rangle$ and $|\psi\rangle$, measured in the basis $M=\{ |i\rangle \langle i|, |j\rangle \langle j| \}$. \\
Then for $|i\rangle$,
\begin{align}
p(\lambda_j, \hat{\lambda}||i\rangle,M) &= 1/(4\pi) \times \Theta( \hat{j} \cdot ( \hat{i} + \hat{\lambda} ))\\
&= 1/(4\pi) \times \Theta( -1 + \hat{j} \cdot\hat{\lambda} ))\\
&= 0 \text{ (almost everywhere)}\\
&\text{and}\\
p(\lambda_i, \hat{\lambda}||i\rangle,M) &= 1/(4\pi) \times \Theta( \hat{i} \cdot ( \hat{i} + \hat{\lambda} ))\\
&= 1/(4\pi) \times \Theta( 1 + \hat{j} \cdot\hat{\lambda} ))\\
&= 1/(4\pi) \text{ (almost everywhere)}
\end{align}
while for $|\psi\rangle$
\begin{align}
p(\lambda_i, \hat{\lambda}||\psi\rangle,M) &= 1/(4\pi) \times \Theta( \hat{i} \cdot ( \hat{\psi} + \hat{\lambda} ))\\
&\text{therefore the overlap is} \nonumber \\
&\int d\hat{\lambda} (1/4\pi) \times \Theta( \hat{i} \cdot ( \hat{\psi} + \hat{\lambda} ))\\
&= |\langle \psi | i \rangle |^2
\end{align}
Thus the Bell-Mermin model, which is $\psi$ ontic, can be modified to MD case to be $\psi$ epistemic.

\chapter{Some observations on Preparation Independence}
PI was introduced in \ref{gandom}. Here we collect a few observations on this assumption.\\

\section{Preparation Independence for product state measurements}\label{prep}
Preparation Independence(PI) was used in \cite{23} as their central assumption. In their paper a product state and entangled measurement basis is used. Here we show that for the case of product state and product state basis, \textit{any} ontological model satisfying PI will satisfy quantum predictions. Thus, PI is a natural assumption for this case.\\ 

\begin{theorem}
An ontological model which correctly reproduces quantum predictions for individual states will reproduce correct predictions for product state measurements on product states if the model satisfies PI.\\
\textbf{Proof}: If we have a product state $ \rho =|\psi\rangle \langle\psi| \otimes|\phi\rangle\langle \phi| $, where both $|\psi\rangle$ and $|\phi\rangle$ are in n-dimensional Hilbert space, on which projective measurement is performed, where the measurement basis is also product state,
\begin{align}
 M &= ( |e_i\rangle \langle e_i|\otimes |f_j\rangle \langle f_j|)_{i,j \in \{1,2,3...n\}} \\
 &=  ( |e_i\rangle \langle e_i|)_{i \in \{1,2,3...n\}} \otimes (|f_j\rangle \langle f_j|)_{j \in \{1,2,3...n\}} \\
&=  \begin{pmatrix} 
|e_1\rangle \langle e_1| \\
|e_2\rangle \langle e_2| \\
.. \\
..\\
|e_n\rangle \langle e_n|
\end{pmatrix} \otimes \begin{pmatrix}
|f_1\rangle \langle f_1| \\
|f_2\rangle \langle f_2| \\
.. \\
..\\
|f_n\rangle \langle f_n|
\end{pmatrix} \\
 &= M_1 \otimes M_2
\end{align} satisfying $ \langle e_i | e_j \rangle = 0 $ and $ \langle f_i | f_j \rangle = 0 $, then probability of getting the $k^{th}$ outcome out of $n^2$ possibilities, corresponding to $ k = j + (i-1)n $, where each k corresponds to a unique i and j:
\begin{align}
p(k | \rho, M) &= tr_{AB}( |\psi\rangle \langle\psi| \otimes |\phi\rangle\langle \phi| \times |e_i\rangle \langle e_i|\otimes |f_j\rangle \langle f_j| ) \\
&= tr_A(|\psi\rangle \langle\psi|e_i\rangle \langle e_i|) \times tr_B(|\phi\rangle \langle\phi|f_j\rangle \langle f_j|) \\
&= \int d\lambda_A p(i | \lambda_A, M_1) p(\lambda_A ||\psi\rangle \langle\psi|) \int d\lambda_B p(j | \lambda_B, M_2) p(\lambda_B ||\phi\rangle \langle\phi|) \\
&= \int \int d\lambda_A d\lambda_B \times p(i | \lambda_A, M_1)p(j | \lambda_B, M_2)\times   p(\lambda_A ||\psi\rangle \langle\psi|)p(\lambda_B ||\phi\rangle \langle\phi|)\\
&= \int \int d\lambda_A d\lambda_B \times p( k | \lambda_A,\lambda_B, M_1\otimes M_2)\times   p(\lambda_A ||\psi\rangle \langle\psi|)p(\lambda_B ||\phi\rangle \langle\phi|)
\end{align}
\end{theorem}
It may be noted that one cannot rule out from this theorem models that do not satisfy PI and still reproduce quantum predictions.\\

\section{Weakening the Preparation Independence Postulate}
Hall gave a different proof of PBR theorem \cite{23} based on weakened assumptions\cite{41}. Instead of Preparation Independence, he formulated "compatibility" and "local compatibility". Here we take a look and see why they are "weaker".\\

The PBR theorem first assumes separability for product states.
\begin{definition}
A hidden variable model is separable if
$ p( \lambda | ( \rho = \rho_1 \otimes \rho_2 \otimes....\otimes  \rho_n), M) > 0$ $\Rightarrow \lambda=(\lambda_1,\lambda_2,....,\lambda_n)$ where $\lambda_i$ is the ontic state of $i^{th}$ system. 
\end{definition}
They further assume that the hidden variable state of each individual system is independent of all others:
\begin{equation}
p(\lambda_1,\lambda_2,....,\lambda_n | \rho_1 \otimes \rho_2 \otimes....\otimes  \rho_n, M ) = \prod_{i=1}^n p(\lambda_i | \rho_i, M)
\end{equation}

\subsection{Compatibility}\label{compat}
Denoting by the condition $\lambda \sim \{ \rho = (|\psi_1\rangle  \langle \psi_1 | \otimes \hat{I} \otimes \hat{I} ....\otimes \hat{I}) , M \}$  only if $p(\lambda | \rho, M ) > 0 $, we proceed to consider the case of 2 qubits in product state as in PBR theorem.\\

\begin{definition}
A hidden variable model is compatible if\\
 $a)$ $\lambda \sim \{ \rho = (|\psi\rangle  \langle \psi | \otimes \hat{I}) , M \} $\\
$b)$ $\lambda \sim \{ \rho = (|\phi\rangle  \langle \phi | \otimes \hat{I}) , M \} $ \\
imply the following\\
$i)$ $\lambda \sim \{ \rho = (|\psi\rangle  \langle \psi| \otimes |\phi\rangle  \langle \phi |), M \}$,\\
$ii)$ $ \lambda \sim \{ \rho = (|\phi\rangle  \langle \phi| \otimes |\psi\rangle  \langle \psi |) , M \}$,\\
$iii)$ $ \lambda \sim \{ \rho = (|\psi\rangle  \langle \psi| \otimes |\psi\rangle  \langle \psi |) , M \}$,\\
$iv)$ $ \lambda \sim \{ \rho = (|\phi\rangle  \langle \phi| \otimes |\phi\rangle  \langle \phi |) ,M \} $ \\
 where $iii)$ follows from $a)$ and $iv)$ follows from $b)$, and $i)$ and $ii)$ follow from $a)$ and $b)$ jointly. \\
\end{definition}
Now let there be a preparation procedure which prepares either of $\rho_{\{1,2,3,4\}}$ = $\{ |\psi\rangle  \langle \psi| \otimes |\phi\rangle  \langle \phi |, |\psi\rangle  \langle \psi| \otimes |\psi\rangle  \langle \psi |, |\phi\rangle  \langle \phi| \otimes |\phi\rangle  \langle \phi |, |\phi\rangle  \langle \phi| \otimes |\psi\rangle  \langle \psi | \}$. 


In the measurement basis $M$ considered in PBR, $p(k | M, \lambda) = 0$ if  $\lambda \sim \{ \rho_k , M \}$. Now assume epistemicity. Let $\exists\text{ } \lambda \sim \{ \rho_k , M \}$ for $ k=1,2,3,4$ over a set S. However $\lambda \in \text{S} \Rightarrow p(k | \lambda, M) = 0$. But we know $\sum_k p(k | \lambda, M) =1 $. So there's a contradiction, and hence S is a null set.\\

Denote by the S$_m$ the set $\{ \lambda \text{ }|\text{ } \lambda \sim \{ \rho = (|\psi\rangle  \langle \psi | \otimes \hat{I}) , M \} \cap \lambda \sim \{ \rho = (|\phi\rangle  \langle \phi | \otimes \hat{I}) , M \}\}$
From Compatibility we know, $ \text{S}_m \subseteq \text{S}$. Hence $ \text{S}_m$ is also a null set. Hence we derive the same conclusion of PBR, without even assuming separability for product states. Thus Compatibility is weaker than Preparation Independence.\\

\subsection{Local Compatibility}
Unlike Compatibility, Local Compatibility assumes separability. Hence it is stronger than Compatibility.\\

\begin{definition}
A hidden variable model is locally compatible if
$ \lambda_i\sim ( \rho_i,  M )$ $\forall i$ $\Rightarrow \lambda=(\lambda_1,\lambda_2,....,\lambda_n) \sim ( \rho_1 \otimes \rho_2 \otimes....\otimes  \rho_n,  M )$.\\
\end{definition}
Considering again the case of 2 qubits,\\
If $ \lambda_1 \sim ( |\psi\rangle  \langle \psi|, M), \text{ }\lambda_2 \sim ( |\phi\rangle  \langle \phi|, M)$ then, it follows from Local Compatibility, $\exists \text{ } \lambda= (\lambda_1,\lambda_2) | \lambda \sim (|\psi\rangle  \langle \psi| \otimes |\phi\rangle  \langle \phi | , M)$.\\

Now consider epistemicity :\\
Let $\exists\text{ } \lambda'\text{ }|\text{ } \lambda' \sim \{ |\phi\rangle  \langle \phi |, M \}, \lambda' \sim \{ |\psi\rangle  \langle \psi |, M \}$.\\

Then from Local Compatibility, we have $ \exists \text{ }\lambda\text=(\lambda',\lambda') \text{ } \text{ }| \text{ } \lambda \sim (\rho_{1,2,3,4}, M)$. But for such $\lambda$ we have already proven a contradiction in previous section \ref{compat}. Thus, either or both of our assumptions, Local Compatibility and epistemicity, must be incorrect. Assuming Local Compatibility as correct, epistemicity is ruled out.\\

It is clear that Local Compatibility is a weaker form of PI, as the former assumes separability but makes no assumption about individual hidden variables being uncorrelated. Compatibility is weaker than both. \\

\chapter{Conclusion}
\section{List of new results}
By violating MI we were led to a class of hidden variable models for Quantum Mechanics that have not been explored in detail yet by researchers in Quantum foundations. Several new MD hidden variable models have been developed in the thesis:\\
1. The $\psi$ ontic deterministic MD model in section\ref{ontic}.\\
2. Modified KS Model I.\ref{modks1}\\
3. Modified KS Model II.\ref{modks2}\\
4. Modified Hall Model.\ref{ryu}\\
5. Modified Bell-Mermin Model.\ref{BM}\\
6. Generalised Brans Model.\ref{omain}\\

We saw, not unsurprisingly, that several theorems valid for ontological models cannot be generalised to MD case in \ref{val}. Some of the theorems however, did retain their validity after appropriately generalising some definitions and hence retain their importance. In particular, none of the theorems that invalidated maximal epistemicity for $d_N \geq 3$ could be generalised, and we later saw that the generalized Brans model is maximally epistemic in arbitrary Hilbert space dimensions. On the other hand, the relation Preparation noncontextual $\Rightarrow$ Maximally $\psi$-epistemic remains true for both ontological and MD models. For these results we generalised the notion of degree of epistemicity to MD case in \ref{definition}.\\

We investigated the correlation between hidden variables and measurement choices that the generalised Brans model exhibits, and noted that there are no cross correlations(Alice's particle correlated with Bob's measurement choice for example) in\ref{new}. We also proved that the model does not satisfy Preparation Independence in\ref{gandom}, and that it has zero randomness despite being reciprocal in \ref{random}.\\

We developed a protocol in \ref{bitchplease} to use Modified KS Model II to simulate quantum channels for qubits with an asymptotic communication cost of 2 bits. The particular advantage of MD models in this case, that they can be used to develop finite communication protocols even in higher dimensions using maximal epistemicity, was noted.\\

The Hall model was modified to develop a model for EPR scenario in \ref{ryu}, which unlike Brans model, has cross correlations between measurement choices and the particles and is still local-causal. This leads to our conclusion that a wrong formulation of local-causality has been used by the authors of \cite{16}. In particular we note that a $\psi$ ontic local model could in principle exist if the correlations be made strong enough.\\

We modified the $\psi$ ontic Bell-Mermin model to MD case in \ref{BM}, which was maximally $\psi$ epistemic. We thus note that the degree of epistemicity of a model is not conserved when measurement dependence is introduced in it.\\

Lastly, it was shown in \ref{prep} that Preparation Independence would always give the correct statistics when the states as well as measurement basis are product states. It is thus a more "natural" assumption for such cases, unlike the case considered in PBR which involves an entangled measurement basis.\\

\section{Future Directions}
PBR theorem, as pointed out by MJW Hall in \cite{41}, was a landmark result in Quantum foundations as it did not use MI to prove its result. We saw the validity of several theorems proven for ontological models for MD case, and an exhaustive list of theorems that can be carried over will give a picture of the structure hidden variable theories take if they are measurement dependent.\\

Regarding simulation of quantum channels using MD models, we saw a two fold increase in communication cost due to Bob having to discard a portion of the information Alice is sending. Is there a protocol that can avoid it is an important question to look into. Also, a protocol to simulate quantum channels in higher dimensions using maximally epistemic MD models is in principle possible, and should be developed.\\

Lastly, we saw that there are models for singlet state correlations that have cross correlations between measurement choices and the particles, and are still local. This opens the possibility of having a $\psi$ ontic MD local model. It is an important question whether such a model actually exists.

\end{document}